\documentclass[12pt, a4paper]{article}
\pdfoutput=1

\usepackage{amsmath}
\usepackage{amsfonts}
\usepackage{amssymb}
\usepackage{graphicx, rotating}
\usepackage{epstopdf}
\usepackage{epsfig}
\usepackage{latexsym}
\usepackage{graphicx}
\usepackage{color}
\usepackage{amsmath,bm,amssymb,bbm}
\usepackage{cite}
\usepackage{slashed}
\usepackage{hyperref}
\usepackage{arydshln}
\hypersetup{colorlinks, citecolor=bluscuro, linkcolor=grigioaltocontrasto,urlcolor=bluscuro,
linktocpage}
\definecolor{rossos}{cmyk}{0,1,1,0.55}
\definecolor{bluscuro}{rgb}{0.15, 0.2, .85}
\definecolor{grigioaltocontrasto}{rgb}{0.37, 0.37, 0.37}
\definecolor{bluchiaro}{cmyk}{1,.3,0.,0.1}

\usepackage[usenames,dvipsnames]{xcolor}

\usepackage{multirow}
\usepackage{enumerate}
\usepackage{cancel}

\newcommand{\hhref}[1]{\href{http://arxiv.org/abs/#1}{arXiv:#1}}

\def\Id{\mathbbm{1}}

\newcommand{\eq}[1]{eq.~(\ref{#1})}

\newcommand{\op}{\mathcal{O}}

\newcommand{\nn}{\nonumber}

\newcommand{\be}{\begin{equation}}
\newcommand{\ee}{\end{equation}}
\newcommand{\bea}{\begin{eqnarray}}
\newcommand{\eea}{\end{eqnarray}}
\newcommand{\bc}{\begin{center}}
\newcommand{\ec}{\end{center}}
\newcommand{\ba}{\begin{array}}
\newcommand{\ea}{\end{array}}

\newcommand{\deltapt}{p_{T,VV}}
\newcommand{\met}{\cancel{E}_T}
\newcommand{\ptvmin}{p_{T,V}^{(min)}}
\newcommand{\thetastar}{\theta^{*}}
\newcommand{\aq}{a_{q}^{(3)}}
\newcommand{\ptv}{p_{T,V}}
 
\newcommand{\TeV}{\,\mathrm{TeV}}

\newcommand{\gev}{\,\mathrm{GeV}}

\def\lra#1{\overset{\text{\scriptsize$\leftrightarrow$}}{#1}}

\begin{document}


\vspace*{-2cm}
\begin{flushright}
CERN-TH-2017-252 \\
RM3-TH/17-1 \\
\end{flushright}

\begin{center}
\vspace*{15mm}

\vspace{0.8cm}
{\bf \Large Electroweak Precision Tests \\ in \\[10pt]  High-Energy Diboson Processes}\\

\vspace{0.9cm}

{\bf Roberto Franceschini$^{a}$,  Giuliano Panico$^{b}$,  Alex Pomarol$^{b,c}$,\\
  Francesco Riva$^d$  and Andrea Wulzer$^{d,e,f}$}

\vspace{.5cm}
\centerline{$^a${\it Dipartimento di Matematica e Fisica, Universit\`a di Roma Tre and INFN, 
00146~Rome}}
\centerline{$^b${\it IFAE and BIST, Universitat Aut\`onoma de Barcelona, 08193~Bellaterra,~Barcelona}}
\centerline{$^c${\it  Dept.~de~F\'isica, Universitat Aut{\`o}noma de Barcelona, 08193~Bellaterra,~Barcelona}}
\centerline{$^d${\it Theoretical Physics Department, CERN, Geneva, Switzerland}}
\centerline{$^e${\it Institut de Th\'eorie des Ph\'enomenes Physiques, EPFL, Lausanne, Switzerland}}
\centerline{$^f${\it Dipartimento di Fisica e Astronomia, Universit\'a di Padova and INFN Padova, Italy}}
\end{center}
\vspace{.5cm}
\begin{abstract}
\medskip
\noindent  A promising avenue to perform precision tests of the SM at the LHC is to measure differential cross-sections at high invariant mass, exploiting in this way the growth with the energy of the corrections induced by heavy new physics. We classify the leading growing-with-energy effects in longitudinal diboson and in associated Higgs production processes, showing that they can be encapsulated in four real ``high-energy primary'' parameters. We assess the reach on these parameters at the LHC and at future hadronic colliders, focusing in particular on the fully leptonic $WZ$ channel that appears particularly promising. The reach is found to be superior to existing constraints by one order of magnitude, 
providing a test of  the SM electroweak  sector  at the per-mille level, 
 in competition  with LEP bounds.
Unlike LHC run-1 bounds, which only apply to new physics effects that are much larger than the SM in the high-energy tail of the distributions, the probe we study applies to a wider class of new physics scenarios where such large departures are not expected.

\end{abstract}

\newpage


\section{Introduction: Energy {\underline{and}} Accuracy}\label{sec:0}

Precision physics is playing an increasingly important role at the LHC. The large luminosity that is being collected will allow for increasingly accurate measurements of SM processes, to be turned into indirect probes of Beyond the SM (BSM) physics. This will require the development of a comprehensive and systematic precision program, analog to the one of ElectroWeak Precision Tests (EWPT) performed at LEP.  The key elements of the LHC precision program are becoming more and more clear. First of all, the lack of direct new particles discoveries suggests that we should focus on heavy new physics, 
at a scale $M$ much above the electroweak (EW) scale. Hence the new physics effects we are searching for are well captured by higher-dimensional operators within an Effective Field Theory (EFT) formalism.  Second, we know that  higher-dimensional operators can be probed both by low energy and by high-energy measurements, and that there is an interplay between the two search strategies. Low energy probes are for instance Higgs coupling measurements, successfully performed already with run-$1$ LHC data \cite{Khachatryan:2016vau}. The advantage of such measurements is that 
they target relatively large (resonantly enhanced) cross-sections. The disadvantage is that they will soon be limited by systematic uncertainties (see e.g.~\cite{CMS:thing} for Higgs couplings), which are unavoidably large at a hadron collider.  On the other hand, high-energy probes are based on the observation that leading-order  higher-dimensional operators can produce, in specific scattering processes, corrections to the high-energy differential cross-section that grow quadratically with the center of mass energy ($E$) relative to the SM prediction. Provided such a growing behavior occurs in a process which one can really measure at high enough energy, new physics effects can become large enough to overcome systematic uncertainties.

The effectiveness of high-energy probes is well understood in the literature, and particularly so in the context of diboson processes \cite{Hagiwara:1986vm,Hagiwara:1989mx,Khachatryan:2015sga,Aad:2016ett,Butter:2016cvz,Zhang:2016zsp,Green:2016trm,Biekoetter:2014jwa,Falkowski:2015jaa,Baglio:2017bfe}. Less understood is the crucial role played by {\emph{accuracy}}, namely the fact that measurements of high energy cross-sections can be turned into  more valid and informative probes of new physics only if they are accurate enough \cite{Farina:2016rws}. Specifically, rough upper bounds on the high energy cross-section in excess to the SM, such as those one could for instance extract from the recasting of resonance searches, would not suffice for our purposes. A dedicated program of accurate measurements is needed. The point is that inaccurate measurements are only sensitive to large (say, order one) relative departures from the SM, hence they can only probe new physics scenarios that foresee such large deviations. In diboson production processes large deviations are possible only in  some exotic strongly-coupled
scenarios \cite{Liu:2016idz}, but they are not generically expected.
\footnote{There are other channels where $O(1)$ departures are instead expected   in very well motivated BSM scenarios. 
Vector bosons scattering is a prominent example.} In fact, in most ``minimal'' BSM scenarios, notably those aimed at addressing the naturalness problem, 
it happens that new physics resonances kick in before the quadratic enhancement makes the BSM contribution to the scattering amplitude larger than the SM one. In general new physics models, BSM particle production occurs at the typical mass scale $M$ of the new physics sector, which acts as the cutoff of the EFT description. Depending on the underlying UV model, the amplitude growth can be smoothly saturated at that scale, or display a resonant peak that one could more effectively see by dedicated resonance searches. In no case it will display the growing with energy behavior predicted by the EFT, making our search strategy ineffective. Accurate experimental measurements that are sensitive to relatively small BSM effects, still performed at high energy such as to exploit the enhancement as much as possible, are needed in order to overcome this potential limitation.

\begin{figure}[t]
\centering
\includegraphics[width=0.7\textwidth]{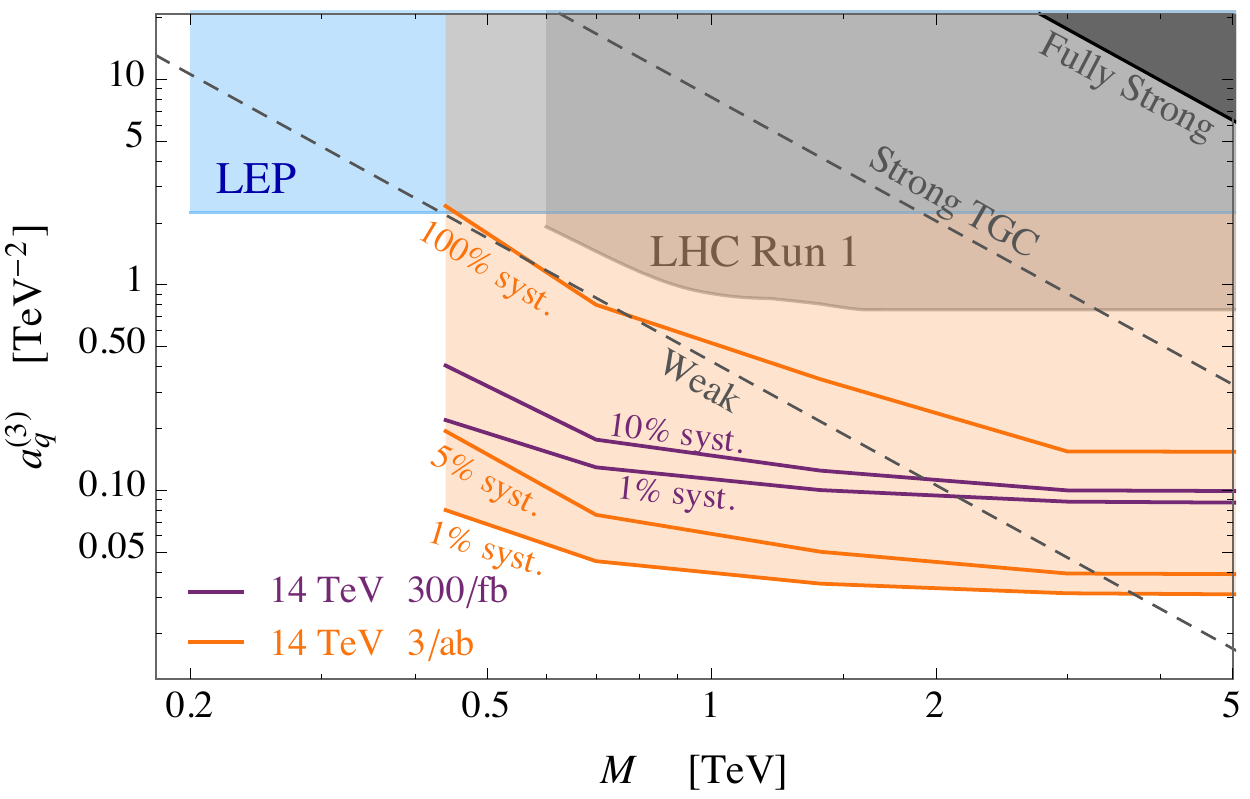}
 \caption{{Bounds from LEP \cite{LEP:2003aa}, run-$1$ LHC (which includes $20\,$fb$^{-1}$ at $8\,$TeV and $3\,$fb$^{-1}$ at $13\,$TeV) \cite{Falkowski:2016cxu}, and the expected $95\%$ CL reach from fully leptonic $WZ$, on the high-energy primary parameter $a^{(3)}_q$ as a function of the new physics scale $M$. See section~\ref{sec:nloanalysis} for a detailed description of the figure.}}
\label{cvsm}
\end{figure}

We can quantitatively illustrate this point by anticipating some of our results, reported in figure~\ref{cvsm}. The figure shows the $95\%$~CL reach, in the $WZ$ production process, on one of our ``high-energy primary'' parameters ($a^{(3)}_q$, introduced in section~\ref{sec:thf}) that describe growing-with-energy effects in the amplitude  for diboson production. In particular, in the $WZ$ channel
\be
\delta {\cal A}(\bar q q^\prime\to WZ)\sim  a^{(3)}_qE^2\,.
\ee
 The reach on $a^{(3)}_q$ is displayed as a function of the cutoff scale $M$, and it is obtained by including in the analysis only events that occur at a center of mass energy $m_{\textsc{wz}}$ below $M$, i.e., events that originate in an energy regime where the EFT prediction is trustable and the energy growth is physical. The different lines correspond to different assumptions about the systematic relative uncertainty in the experimental measurement of the differential cross-section and in the theoretical prediction of the SM contribution. The ``$\delta_{\rm{syst}}=100\%$'' curve corresponds to an inaccurate determination of the cross-section, which is only sensitive to  order one departures from the SM.  In the figure, the reach on $a^{(3)}_q$ is compared with theoretical expectations on the relation between $a^{(3)}_q$ and $M$. The  line  ``Fully Strong" corresponds to the rather implausible (although, strictly speaking, allowed)  physical situation   where all the particles involved in the scattering (i.e., the bosons and the light quarks) have  maximal  couplings, $\sim 4\pi$, to the new physics sector at the scale $M$.  This line is then given by $a^{(3)}_q=16\pi^2/M^2$, and the dark region above it is excluded by perturbative unitarity. The  line ``Weak'' corresponds to $a^{(3)}_q=g^2/M^2$,  where $g$ is the SU$(2)_L$ SM coupling, and it is around this line where the most interesting BSM scenarios live. These are scenarios where the SM gauge bosons and the light quarks are ``elementary'', i.e. they are coupled only through gauge interactions to the BSM particles  at the scale $M$.\footnote{Other SM particles, such as the Higgs,  could very well be ``composite'', i.e. strongly coupled, in these scenarios. Composite Higgs models are indeed examples of theories that lie around the ``Weak'' line.} In these cases, the BSM amplitude is always smaller than the SM one (which is of order $g^2$) in the whole range of validity of the EFT $E\lesssim M$. Therefore the BSM corrections to the cross-section never overcome the SM expectation, and we cannot probe these scenarios  through inaccurate measurements, as the $\delta_{\rm{syst}}=100\%$ curve shows. The ``Strong TGC'' line (TGC standing for Triple Gauge Couplings) is $a^{(3)}_q=4\pi g /M^2$, and it corresponds to a limiting case of the ``Remedios'' scenario of Ref.~\cite{Liu:2016idz}, where the quarks are elementary while the transverse gauge fields are strongly interacting and partially composite. Notice that there are also many interesting scenarios, such as  supersymmetric theories, where the contributions to $a^{(3)}_q$ arise at the one-loop level, $a^{(3)}_q\sim g^2/(16\pi^2M^2)$,  predicting a line in the plane, not shown in the figure, much below the ``Weak'' one. None of the indirect bounds we are discussing applies to these BSM scenarios. We also report in figure~\ref{cvsm} the LEP and LHC run-$1$ bounds on $a^{(3)}_q$. The former is a horizontal line because it is obtained from low-energy measurements, specifically, from the LEP2 measurement of the $\delta g_1^Z$ anomalous triple gauge coupling at $E\sim 200$ GeV \cite{LEP:2003aa}. The LHC run-$1$ line is derived in Ref.~\cite{Falkowski:2016cxu} from a recasting of the LHC $WW$ and $WZ$ results, considering only signal events with invariant mass below $M$. 

If all the events are used, i.e. for $M\rightarrow\infty$,  figure~\ref{cvsm}  confirms the well-known result that LHC run-$1$ has a better reach than LEP.  Nevertheless, when looking at the full $a^{(3)}_q$-$M$ plane of figure~\ref{cvsm}, we see that the LHC run-$1$  limit only applies to ``Fully Strong'' and to ``Strong TGC''  scenarios, and hence it does not improve LEP in the exploration of  ``Weak'' BSM theories, which on the other hand are the most interesting ones.  This is mainly due to the relatively low energy of run-$1$ collisions, which entails a low rate for high-energy processes and consequently an inaccurate determination (or even actually a mere upper bound) of the cross-section. \mbox{Run-$2$} and run-$3$ data will not suffer from this issue and they will be capable to probe  ``Weak'' theories if accurate enough measurements are performed. A {\emph{qualitative}} improvement in BSM physics exploration, as opposite to a mere quantitative increase of the sensitivity, will thus be possible. Further progress could be made at the High Luminosity (HL) LHC, as the figure shows.

The purpose of this paper is to provide  high-energy probes  for new physics  that can apply to a wide class of BSM theories, 
in special those of the  "Weak" type. 
For this reason we study quadratically growing with energy effects in diboson production processes  that can
arise from dimension-six ($d=6$) EFT operators.
 Since     contributions to the amplitudes  from BSM  of the "Weak"  type  
 are   smaller than the SM ones, as previously explained, sizable $E^2$-enhanced contributions to the differential cross-section are only possible   in the presence of interference between the SM and the BSM terms. For diboson differential cross-section measurements that are inclusive over the bosons decay products, interference emerges only in the production of longitudinally polarized vector bosons (see \cite{Panico:2017??} and references therein). 
Fortunately, these are also the most motivated channels from the BSM perspective mentioned above. The high-energy dynamics of longitudinally-polarized vectors is inextricably linked to the one of the Higgs particle, due to the Equivalence Theorem and  the ${\textrm{SU}}(2)_L\times{\textrm{U}}(1)_Y$ invariance restored at high energies. We will thus have to enlarge the scope of our analysis to include among ``diboson'' processes also $Wh$ and $Zh$ associated Higgs production.

The paper is organized as follows. We first classify and parametrize growing-with-energy effects based on symmetries, with an approach that is independent of the EFT operator basis and in some respect more general than (although in practice equivalent to) the EFT one. We will see that these effect can be encapsulated in four real parameters that we call ``high-energy primaries''. They are the high-energy analog of the primaries defined in Ref.~\cite{Gupta:2014rxa}, which were   instead optimized for parametrizing low-energy effects. High-energy primaries are useful because they offer a concise picture of which effect it is worth looking for in each final state and they outline model-independent connections among different final states. In the perspective of a global fit to the  EFT parameters, which is the final aim of the LHC precision program, synthetic and basis-independent parametrizations of this sort are of utmost importance. In section~\ref{sec:thf} we define the high-energy primaries through the above-mentioned classification, and we illustrate their connection with popular EFT operator bases and with the low-energy primaries. We also describe their origin and expected magnitude in explicit BSM scenarios. Section~\ref{sec:sens} is devoted to LHC phenomenology. We first present a broad overview of diboson channels and a semi-quantitative estimate of the reach. We identify fully leptonic $WZ$ as a promising channel, which we investigate in detail in section~\ref{sec:lWZ}. The implications of the results are discussed in section~\ref{sec:results}. Conclusions and outlook are reported in section~\ref{sec:conc}.

\section{Theoretical Framework}\label{sec:thf}

We are interested in processes which fulfill two conditions. First, their amplitudes must receive BSM contributions that grow with $E^2$ at the leading order (i.e., $d=6$) in the EFT operator expansion.\footnote{We consider here large center of mass energy and large scattering angles, namely large Mandelstam variables $s\sim t\sim u \sim E^2\gg m^2_{W}$.} Second, the  SM amplitudes must be constant and sizable at high energy, in such a way that,
at the linear order in the EFT Wilson coefficient, 
 the $E^2$-growth of the BSM amplitudes results into a $E^2$-growth of the differential cross-sections 
thanks to the SM-BSM interference. 
This condition is required by the fact that we are interested in probing theories whose indirect effects remain smaller than the SM even at high-energy, as previously explained. In table~\ref{energygrowth} we summarize  the high-energy behavior of amplitudes with different diboson helicity configurations, in the SM and in generic BSM (meaning the maximal effect that can be achieved with an insertion of any $d=6$ operator --see for example Ref.~\cite{Azatov:2016sqh}). We focus on same-chirality (i.e., opposite helicity) quark anti-quark initial states because opposite chirality amplitudes are suppressed by the quark Yukawa couplings in the SM, making the interference term negligible in these channels.

 \begin{table}[t]
\begin{center}
\begin{tabular}{c|c|c}
&SM&BSM\\\hline
\rule[-.5em]{0pt}{1.65em}$ q_{L,R}\bar q_{L,R}\to V_L V_L (h)$& $\sim 1$& $\sim E^2/M^2$ \\\hline
\rule[-.5em]{0pt}{1.65em}$ q_{L,R}\bar q_{L,R}\to V_\pm V_L (h)$& $\sim m_W/E$& $\sim m_WE/M^2$ \\\hline
\rule[-.5em]{0pt}{1.65em}$ q_{L,R}\bar q_{L,R}\to V_\pm V_\pm$& $\sim m_W^2/E^2$& $\sim E^2/M^2$ \\\hline
\rule[-.5em]{0pt}{1.65em}$ q_{L,R}\bar q_{L,R}\to V_\pm V_\mp$& $\sim 1$& $\sim 1$ 
 \end{tabular}
  \caption{{High-energy  scaling of diboson amplitudes for transverse ($\pm$)  and longitudinal ($L$) polarizations in the SM and in BSM (parametrized by $d=6$ operators suppressed by $1/M^2$).}}
 \label{energygrowth}
\end{center}
\end{table}

The results of the table can be understood as follows. Maximal helicity violating (MHV) amplitudes $q\bar q\to V_\pm V_\pm$ are suppressed in the SM massless limit \cite{Parke:1986gb,Berends:1997js}, and scale like $m^2_W/E^2$ for finite mass; MHV selection rules don't apply in BSM, where they grow therefore unsuppressed. On the other hand, $q\bar q\to V_\pm V_\mp$ are not suppressed in the SM at  high-energy, but don't receive contributions from $d=6$ operators \cite{Simmons:1989zs,Azatov:2016sqh}. The suppression of SM amplitudes with one longitudinal only can be understood as a consequence of the symmetry under which all the SM doublets (Higgs and fermions) change sign, namely $H\to-H$, $Q_L\to -Q_L$ and $L_L\to -L_L$. This operation corresponds to the ``$g_L=-\Id$'' element of SU$(2)_L$, which is part of the SM gauge group and hence it is respected both by the SM and the BSM Lagrangian. Since the symmetry is only broken by the Higgs VEV $v$, it produces a selection rule that controls whether even or odd powers of $v$ (actually, of $m_W$) are present in the amplitudes \cite{Borel:2012by}. Transversely polarized vector bosons are even under the symmetry, while longitudinal polarizations are odd because they are related to the Goldstone components of the Higgs doublet through the Equivalence Theorem.\footnote{The fact that longitudinals are odd can be established also in the unitary gauge, by noticing that the longitudinal polarization vectors are proportional to $1/m_{W,Z}$.} The amplitudes for producing one transverse and one longitudinal state (or a Higgs) are odd, hence they scale like $m_W/E$ and $m_WE/M^2$, respectively,  in the SM and in the $d=6$ EFT, as the table shows.

In summary, we see that $V_LV_L $ and $V_L h$ production are the only processes that display quadratic energy growth at the interference level; we thus focus on these in the rest of the paper. Notice however that promising strategies to circumvent the non-interference problem have been recently proposed \cite{Panico:2017??,Azatov:2017kzw}, which allow for instance to ``resurrect'' interference effects in transverse vector bosons production. Since these strategies require measuring additional observables other than the diboson differential cross-sections that we consider here, we leave to future work studies in this direction.

\subsection{High-Energy Primary Effects}
 
The study of longitudinally-polarized dibosons production in the high-energy limit $E\gg m_{W}$ is greatly simplified by using the Equivalence Theorem~\cite{Chanowitz:1985hj}, and its more systematic formulation in Ref.~\cite{Wulzer:2013mza}. In this formalism, external longitudinally-polarized vector states are represented in Feynman diagrams as the corresponding scalar Goldstone bosons, up to corrections of order $m_W/E$ from diagrams with gauge external lines. Furthermore, the $E\gg m_W$ limit can be safely taken in the internal line propagators and in the vertices, making that all the effects (masses and vertices) induced by the Higgs VEV manifestly produce order $m_W/E$ corrections. In order to assess the leading energy behavior, it is thus sufficient to study the amplitude in the unbroken phase, where the EW bosons are massless and the $G_{\rm{SM}}={\textrm{SU}}(2)_L\times{\textrm{U}}(1)_Y$ symmetry is exact. Given that the Goldstone bosons live in the Higgs doublet $H$, together with the Higgs particle, $G_{\rm{SM}}$ implies that the high-energy behavior of the former ones are connected with the latter. This is the technical reason why $V_LV_L$ and $V_Lh$ production processes, collectively denoted as $\Phi\Phi'$ in what follows, should be considered together, like we do in the present article.

\begin{figure}[t]
\centering
\includegraphics[width=.9\textwidth]{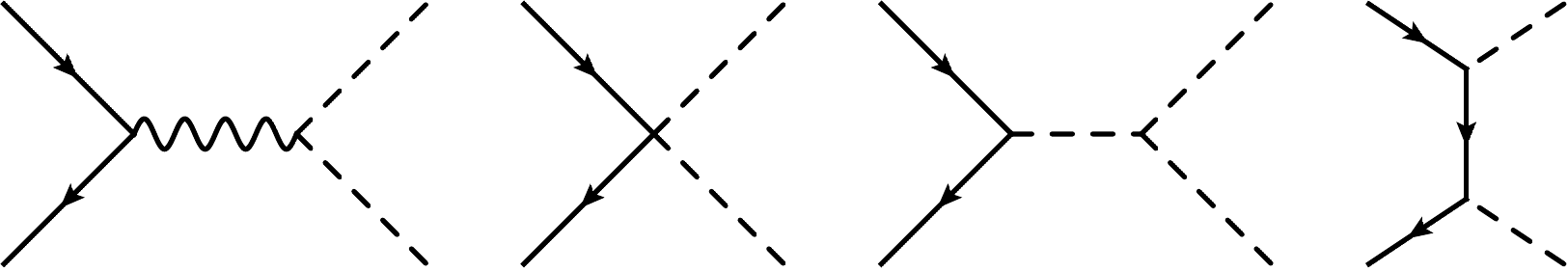}
 \caption{{Representative diagrams for $q'{\bar{q}}\rightarrow\Phi\Phi'$ production.}}
\label{diag}
\end{figure}

We consider the production of $\Phi\Phi'$ out of a quark $q'$ with helicity $\lambda'$ and an anti-quark ${\overline{q}}$ with helicity $\lambda$, with the aim of classifying possible growing-with-energy contributions induced by higher-dimensional operators, in particular those that do interfere with the SM.  The tree-level Feynman diagrams responsible for the process, schematically depicted in fig.~\ref{diag}, can have $s$-channel, $t$-(or $u$-)channel, or contact interaction topology. The $s$-channel gauge bosons exchange (first diagram) is the only relevant topology in the SM because $\Phi$ vertices with the light quarks are proportional to the tiny Yukawa couplings. In the SM, the process thus exclusively occurs in the $J=1$ angular momentum configuration. Furthermore, because of the structure of the fermion-gauge-boson vertex, it is necessarily initiated by quarks and anti-quarks with opposite helicity, i.e. $\lambda\neq\lambda'$. All the quark flavor combinations are possible in the SM, aside from $u_+{\overline{d}}_-$ and $d_+{\overline{u}}_-$  that vanish in the SM due to the absence of $W$ couplings to right-handed quarks. BSM effects that interfere with the SM must thus also occur in opposite-helicity quark anti-quark scattering, with the exception of $u_+{\overline{d}}_-$ and $d_+{\overline{u}}_-$.

We are interested here in the leading order effects in an EFT expansion, which are of order $E^2/M^2$ by dimensional analysis. 
These effects can emerge from the insertion of one anomalous vertex in the $s$- or $t$-channel diagrams, or from contact interactions. Among the former diagrams, $s$-channel gauge bosons exchange is once again the only relevant topology because the others require one insertion of the SM Yukawa couplings. These contribute to the $J=1$ angular momentum configuration like the SM terms. Contact interaction terms can in principle contribute to all partial waves, however it is not hard to see that only $J=1$ is possible for dimension-six operators. This follows from the fact that $J\geq2$ would require more derivative/fields than those allowed by dimensionality and that $J=0$ $\Phi\Phi'$ production from opposite-helicity quark and anti-quark would require operators with one right-handed fermion singlet, one left-handed fermion doublet and two Higgs doublets that are forbidden by the SM group. In conclusion, the relevant BSM effects can be parametrized as corrections to the $J=1$ partial wave amplitudes, namely
\begin{equation}\label{amp0}
\delta{\mathcal{A}}\left(q^\prime_{\pm}{\overline{q}}_{\mp}\rightarrow\Phi\Phi'\right)= f^{\Phi\Phi'}_{q^\prime_{\pm}{\overline{q}}_{\mp}}(s)\sin\theta=  \frac{1}{4} A^{\Phi\Phi'}_{q^\prime_{\pm}{\overline{q}}_{\mp}}\, E^2 \sin\theta^*\,,
\end{equation}
where $\theta^*$ is the scattering angle in the center of mass, and $E=\sqrt{s}$ is the center of mass energy. The azimuthal angle, upon which the amplitude depends as $e^{\pm i\phi}$, has been set to zero for shortness. The dependence on $\theta^*$ (and on $\phi$) is fixed by angular momentum conservation, as a simple application of the Jacob-Wick formula \cite{Jacob:1959at} to the case $J=1$, $\lambda_{in,1}-\lambda_{in,2}=\pm1$ and $\lambda_{fin,1}-\lambda_{fin,2}=0$. 

Notice that the contact interaction topology of fig.~\ref{diag} can a priori produce BSM effects with a non-trivial structure in the quark family space. However flavor physics tightly constraints \cite{Bauer:2009cf,Buras:2011ph,Panico:2016ull} non-universal contact interactions involving the light generations, which are the only ones that are relevant for the LHC diboson production.  We can thus assume flavor universality without loss of generality.

Eq.~(\ref{amp0}) shows that at the leading order in the SM EFT expansion each diboson process is sensitive at high energy to a single constant new-physics parameter $A^{\Phi\Phi'}_{q^\prime_{\pm}{\overline{q}}_{\mp}}$ for every combination of initial or final states. This can be taken real since its imaginary part does not interfere with the SM. In addition, the SM symmetry group, which is restored in the high-energy limit, as previously explained, implies  several relations among these parameters. Namely \footnote{Below and in what follows we work for simplicity with diagonal Cabibbo--Kobayashi--Maskawa (CKM) matrix. Otherwise the relations that follow hold in the quark interaction basis and need to be rotated to the mass basis, producing CKM factors in the charged amplitudes.} 
\begin{eqnarray}\label{primaries0}
&A^{W^+W^-}_{u_+{\overline{u}}_-}=A^{Zh}_{u_+{\overline{u}}_-}=-a_u\,,\;\;\;\;\;A^{W^+W^-}_{d_+{\overline{d}}_-}=A^{Zh}_{d_+{\overline{d}}_-}=-a_d\,,&\nonumber\\
&A^{W^+W^-}_{u_-{\overline{u}}_+}=A^{Zh}_{d_-{\overline{d}}_+}=a_q^{(1)}+a_q^{(3)}\,,\;\;\;\;\; A^{W^+W^-}_{d_-{\overline{d}}_+}=A^{Zh}_{u_-{\overline{u}}_+}=a_q^{(1)}-a_q^{(3)}\,&\nonumber\\
&A^{hW^+}_{u_-{\overline{d}}_+}=A^{ZW^+}_{u_-{\overline{d}}_+}=A^{hW^-}_{d_-{\overline{u}}_+}=-A^{ZW^-}_{d_-{\overline{u}}_+}=\sqrt{2}a_q^{(3)}\,&
\label{relation}
\end{eqnarray}
where $a_u$, $a_d$, $a_q^{(1)}$ and $a_q^{(3)}$ are the coefficients of the decomposition of the amplitude in $G_{\rm{SM}}$-invariant tensors, which we work out in Appendix~\ref{appA}. In $a_u$, $a_d$ and $a_q^{(1)}$ the incoming (and outgoing) states form an $\textrm{SU}(2)_L$ singlet, while in $a_q^{(3)}$ they form a triplet. The four quantities  $a_u$, $a_d$, $a_q^{(1)}$ and $a_q^{(3)}$ define our high-energy primaries (HEPs). They parametrize  all possible BSM effects that produce quadratic energy growth at the interference level in diboson production at high-energy, as summarized in the first two columns of table~\ref{Wilsons}. Notice that the HEP parameters have energy dimension $-2$; we will measure them in units of  TeV$^{-2}$ in what follows. 

The fact that only the $4$ HEP parameters produce sizable effects at high energy is non-trivial from the point of view of the generic $d=6$ EFT, where a total of $6$ anomalous couplings contribute to longitudinal diboson processes. These couplings can be identified as
${\delta g^Z_{uL}}$, ${\delta g^Z_{uR}}$, ${\delta g^Z_{dL}}$, ${\delta g^Z_{dR}}$, ${\delta g_1^Z}$ and ${\delta \kappa_{\gamma}}$ in the notation of Ref.~\cite{Gupta:2014rxa}, defined through their contributions to trilinear vertices as
\bea
\Delta{\cal L}_{\rm BSM}&=&
{\delta g^Z_{uL}}\left[Z^\mu \bar u_L\gamma_\mu u_L+\frac{c_{\theta_W}}{\sqrt{2}}(W^{+\mu}\bar{u_L}\gamma_\mu d_L +\textrm{h.c.})+\cdots \right]
+{\delta g^Z_{uR}}\,\left[ Z^\mu \bar u_R\gamma_\mu u_R+\cdots \right]\nn\\
&+&
{\delta g^Z_{dL}} \left[Z^\mu \bar d_L\gamma_\mu d_L-\frac{c_{\theta_W}}{\sqrt{2}}(W^{+\mu}\bar{u}_L\gamma_\mu d_L +\textrm{h.c.}) +\cdots \right]
+{\delta g^Z_{dR}}\, \left[ Z^\mu \bar d_R\gamma_\mu d_R+\cdots \right]
\nn\\
&+&ig c_{\theta_W} {\delta g_1^Z}\Big[  (Z^\mu (W^{+ \nu} W^{-}_{\mu\nu}-\textrm{h.c.})+Z^{\mu\nu}W^+_\mu W^-_\nu +\cdots \Big]\nn\\
&+&ie\, {\delta \kappa_{\gamma}}\,\left[(A_{\mu\nu}- t_{\theta_W}Z_{\mu\nu})W^{+\mu} W^{-\nu}+\cdots \right]
\,,
\label{primaries}
\eea
where $Z_{\mu \nu}\equiv \hat{Z}_{\mu \nu}-i   W^+_{[\mu}W^-_{\nu]} $, $A_{\mu \nu}\equiv \hat{A}_{\mu \nu}$,  $ W^{\pm}_{\mu\nu}\equiv\hat{W}^{\pm}_{\mu \nu}\pm i  W^{\pm}_{[\mu}(A+Z)_{\nu]} $ with $\hat{V}_{\mu\nu}\equiv\partial_\mu V_\nu-\partial_\nu V_\mu$, and $c_{\theta_W}\equiv \cos\theta_\textsc{w}$ where $\theta_\textsc{w}$ is the weak mixing angle.  Modifications of the left-handed quark couplings to the $W$ are related to modifications to the $Z$ couplings, due to an accidental custodial symmetry present in the dimension-six operators. Similarly, the above $6$ low-energy primary parameters are  related to 
certainx modifications of the physical  Higgs couplings, denoted with dots in \eq{primaries}  (see Ref.~\cite{Gupta:2014rxa} for details). The relations between the HEP parameters and the $4$ combinations of the low-energy primaries that produce growing-with-energy effects are reported in the third column of table~\ref{Wilsons}.

\begin{table}[t]
\begin{center}
\begin{tabular}{c|c|c}
Amplitude& High-energy primaries& Low-energy primaries  \\\hline
\rule[-1.4em]{0pt}{3.2em}$\bar u_L d_L\to W_LZ_L,W_Lh$ & $\sqrt{2}a_q^{(3)}$ & $ \displaystyle\sqrt{2}\frac{g^2}{m_W^2}\left[c_{\theta_W}({\delta g^Z_{uL}}-{\delta g^Z_{dL}})/g-c_{\theta_W}^2{\delta g_1^Z} \right]$ \\
\hline
\rule[-.6em]{0pt}{1.7em}$\bar u_L u_L\to W_LW_L$& \multirow{ 2}{*}{$a_q^{(1)}+a_q^{(3)}$}& \multirow{ 2}{*}{$\displaystyle-\frac{2g^2}{m_W^2}\left[Y_L t^2_{\theta_W}{\delta\kappa_\gamma}+T_Z^{u_L}{\delta g_1^Z}+c_{\theta_W}{\delta g^Z_{dL}}
 /g\right]$}\\
\rule[-.55em]{0pt}{1.45em}$\bar d_L d_L\to Z_Lh$& &\\
\hline
\rule[-.6em]{0pt}{1.7em}$\bar d_L d_L\to W_LW_L$& \multirow{ 2}{*}{$a_q^{(1)}-a_q^{(3)}$}& \multirow{ 2}{*}{$\displaystyle-\frac{2g^2}{m_W^2}\left[Y_L t^2_{\theta_W}{\delta\kappa_\gamma}+T_Z^{d_L}{\delta g_1^Z}+c_{\theta_W}{\delta g^Z_{uL}}
/g\right]$}\\
\rule[-.55em]{0pt}{1.45em}$\bar u_L u_L\to Z_Lh$& & \\
\hline
\rule[-1.2em]{0pt}{3.em}$\bar f_R f_R\to W_LW_L,Z_Lh$& $a_{f}$& $\displaystyle-\frac{2g^2}{m_W^2}\left[Y_{f_R} t^2_{\theta_W}{\delta\kappa_\gamma}+T_Z^{f_R}{\delta g_1^Z}+c_{\theta_W}{\delta g^Z_{fR}}/g\right]$
 \end{tabular}
  \caption{Parameter combinations (in the high- and in the low-energy primary bases) that control $E^2$-enhanced effects in each polarized longitudinal diboson production process. Here, $T_Z^f=T_3^f-Q_fs^2_{\theta_W}$ and $Y_{L,f_R}$ is the hypercharge of the left-handed and right-handed quark (e.g., $Y_L=1/6$).}
\label{Wilsons}
\end{center}
\end{table}

\subsection{BSM Perspective and Connection with EFT}\label{sec:BSMperspective}
\label{bsmsection}

\begin{figure}[t]
\centering
\includegraphics[width=.9\textwidth]{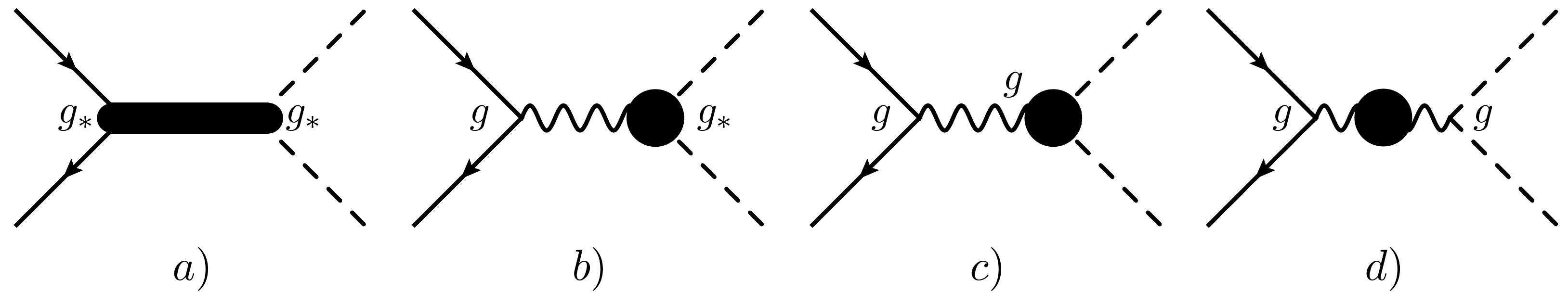}
 \caption{Contributions to longitudinal diboson processes from different BSM scenarios: Strongly-coupled quarks and Higgs (a),
strongly-coupled  Higgs and transverse vectors (b),
and  "Weak" type models (c,d).}
\label{diagsBSM}
\end{figure}

The HEP parameters, denoted collectively by $a$ in what follows, can be thought as a new class of BSM ``Fermi constants''. Explicit BSM models generate HEPs, whose magnitude scales as  $a\sim (coupling)^2/M^2$. As we have seen in the introduction, the actual product of couplings entering this relation depends on the particular BSM scenario we have in mind. We now discuss this aspect in more detail.

In BSM scenarios where some or all the SM particles are strongly coupled to the new dynamics (for instance because they are composite objects), the relevant couplings can be large. This implies that the relative departures from the SM, which are roughly controlled by ${\cal A}_{\textrm{BSM}}/{\cal A}_{\textrm{SM}}\sim a\, E^2/g^2\sim \left(coupling/g\right)^2\left(E/M\right)^2 $, can be larger than one, even for $E\ll M$. The coexistence of the weakly coupled  SM with  
a strongly-coupled BSM at the scale $M$, can be natural if we postulate the presence of approximate global symmetries in the BSM sector, weakly broken by the  SM couplings.   Explicit examples include models of  fermions compositeness (standard \cite{Kaplan:1991dc} or pseudo-Goldstini \cite{Liu:2016idz,Bellazzini:2017bkb}), or models where the gauge bosons have strong multipolar interactions (called \emph{Remedios}) \cite{Liu:2016idz}. 

Among these classes, models where both fermions and  the Higgs are strongly coupled generate large HEP, $a\sim g_*^2/M^2$
(illustrated in figure~\ref{diagsBSM}$a$), where $g_*>g$ is the coupling associated with the new dynamics. If $g_*$ is maximal, $g_*\sim4\pi$, we obtain the scenario denoted ``Fully Strong'' in the introduction. 
Such a scenario, where light quarks are strongly coupled, is however of limited interest in light of strong constraints on light-quark compositeness from di-jet measurements \cite{Alioli:2017jdo,Domenech:2012ai,Bellazzini:2017bkb}. 

In \emph{Remedios} models \cite{Liu:2016idz}, the transverse polarizations of the SM gauge bosons can have strong interactions, generating  large Wilson coefficients in operators involving the field-strengths $W_{\mu\nu}$. If the Higgs is also part of the strongly-interacting sector, one finds $a\sim g g_*/M^2$ (see diagram \ref{diagsBSM}b). For $g_*=4\pi$ this produces the ``Strong TGCs'' case discussed in the introduction. While structurally interesting, it must be appreciated that these scenarios have been designed explicitly to obtain large anomalous TGCs (aTGCs) with  no other purpose.

On the other hand, in a  larger class of BSM scenarios, denoted ``Weak'' in the introduction, SM fermions and gauge bosons are weakly coupled above $M$ (for instance because they are elementary states). Those include many models that solve the hierarchy problem (eg.~composite Higgs models, extra dimensional models, little Higgs, twin Higgs) and are therefore generally better motivated. 
In these models the contributions to HEPs are always mediated by    SM gauge bosons whose coupling is $g$
(see diagrams \ref{diagsBSM}c and  \ref{diagsBSM}d),
and therefore we expect $a\sim g^2/M^2$.

In several new physics scenarios of the "Weak" class, the light SM fermions have negligible direct couplings with the new dynamics, which only interacts with the SM vector and Higgs bosons. These BSM scenarios, that we call "universal",
 are conveniently parametrized at low-energy in the SILH basis~\cite{Giudice:2007fh},\footnote{Our convention is: $ D_\nu   H = \left( \partial_{\nu} -{1 \over 2} i g' B_{\nu} - {1 \over 2}i g \sigma^{a} W^{a}_{\nu}\right) H $, and  $ W_{\mu\nu}^{a}= \partial_{\mu}W_{\nu}^{a}-\partial_{\nu}W_{\mu}^{a}+g\epsilon_{abc}W_{\mu}^{b}W^{c}_{\nu}$, where $\sigma^{(2)}_{12}=-i $, and $\epsilon_{123}=1$.}  where $d=6$ operators are written as a function  of SM  bosons  only~(see table \ref{silhoperators}). The relations between the HEP and the Wilson coefficients in the SILH basis are given by
\be
 a_q^{(3)}=\frac{g^2}{M^2}(c_W+c_{HW}-c_{2W})\ ,\ \  \ \
  a_q^{(1)}=\frac{g'^2}{3M^2}(c_B+c_{HB}-c_{2B})\,, \label{a2c}
\ee
and
\be
a_u=-2a_d=4 a_q^{(1)}\,.
\label{unifersal}
\ee
These relations can also be written using  the $\hat S$, $\hat T$, $W$ and $Y$ parameters (we follow the notation of Ref.~\cite{Barbieri:2004qk}) in addition to the two anomalous  triple gauge couplings (aTGC), $\delta g_1^Z$ and  $\delta \kappa_\gamma$ defined in \eq{primaries}. We have
\be
 a_q^{(3)}=-\frac{g^2}{m_W^2}\left(c^2_{\theta_W}\delta g_1^Z+W\right)\ ,\ \  \ \
  a_q^{(1)}=\frac{g'^2}{3m_W^2}\left(\hat S-\delta\kappa_\gamma+c^2_{\theta_W} \delta g_1^Z-Y\right)\,,
\label{heptosilh}  
\ee
which  can  be useful in order to  compare HEP analyses  from  LHC  with  other experiments, such as LEP. 

 \begin{table}[t]
\begin{center}
\begin{tabular}{l||l}
SILH Basis&Warsaw Basis\\\hline
\rule[-1.2em]{0pt}{3em}$\displaystyle{\cal O}_W=\frac{ig}{2}\left( H^\dagger  \sigma^a \lra {D^\mu} H \right )D^\nu  W_{\mu \nu}^a$&$\displaystyle{\cal O}^{(3)}_L= (\bar{Q}_L \sigma^a\gamma^\mu Q_L)(i H^\dagger \sigma^a\lra D_\mu   H)$\\
\rule[-1.2em]{0pt}{3em}$\displaystyle{\cal O}_B=\frac{ig'}{2}\left( H^\dagger  \lra {D^\mu} H \right )\partial^\nu  B_{\mu \nu}$&$\displaystyle{\cal O}_L=  (\bar{Q}_L \gamma^\mu Q_L)(i H^\dagger \lra D_\mu   H)$\\
\hline
\rule[-.6em]{0pt}{1.5em}$\displaystyle{\cal O}_{HW}=i g(D^\mu H)^\dagger\sigma^a(D^\nu H)W^a_{\mu\nu}$&
$\displaystyle{\cal O}^u_R= (\bar{u}_R \gamma^\mu u_R)(i H^\dagger \lra D_\mu   H)$\\
\rule[-.6em]{0pt}{1.5em}$\displaystyle{\cal O}_{HB}=i g'(D^\mu H)^\dagger(D^\nu H)B_{\mu\nu}$&$\displaystyle{\cal O}^d_R=  (\bar{d}_R \gamma^\mu d_R)(i H^\dagger \lra D_\mu   H)$\\\hline
\rule[-1.2em]{0pt}{3em}$\displaystyle{\cal O}_{2W}=-\frac{1}{2}  ( D^\mu  W_{\mu \nu}^a)^2$&\\
\rule{0pt}{1.5em}$\displaystyle{\cal O}_{2B}=-\frac{1}{2}( \partial^\mu  B_{\mu \nu})^2$&
 \end{tabular}
  \caption{Dimension-six operators  relevant for the high-energy longitudinal diboson production $q\bar q\to W_LV_L,V_Lh$ that interfere with the SM, in the SILH basis \cite{Giudice:2007fh} (left) and in the Warsaw basis \cite{Grzadkowski:2010es} (right). We will use the Wilson coefficient normalization ${\cal L}_{6}=\sum\limits_i c_i{\cal O}_i/M^2$.
\label{warsawoperators}\label{silhoperators}}
\end{center}
\end{table}

It can be instructive to provide a concrete example of this type of models, and the explicit  values of the HEP parameters that are generated. For this purpose,  let us consider  holographic models of composite Higgs \cite{Agashe:2004rs}. One finds  \cite{Giudice:2007fh}, after integrating out the heavy resonances of the model at tree-level:
\be
c_W=c_B=\frac{27\pi^2}{256}\simeq 1.0\ , \ \ \  
c_{HW,HB}= 0\ , \ \ \  
c_{2B,2W}\simeq \frac{g^2}{g^2_*}\ll 1\, ,
\label{precompo}
\ee
where $g_*$ is here the coupling of the composite heavy vectors, and the new-physics scale  $M$ is  identified with the lightest vector-resonance mass. The relation  $c_W= c_B$ in \eq{precompo} is due  to  a   global $O(4)$ symmetry of the model, and $c_{HW,HB}\ll  c_{W,B}$ is a generic consequence of the ``minimal coupling'' hypothesis \cite{Giudice:2007fh,Liu:2016idz}, which is realized not only in holographic models, but also in little Higgs or other weakly-coupled scenarios. Eq.~(\ref{precompo})  leads to the  following predictions:
\be\label{compopred}
a_{q}^{(3)}= \frac{3g^2}{ g'^2}a_{q}^{(1)}\simeq \frac{g^2}{M^2}\ ,\ \ \
a_{q}^{(3)}{m^2_W}= -g^2 c^2_{\theta_W}\delta g_1^Z=\frac{g^2}{2}\hat S
\ ,\ \ \
\delta\kappa_\gamma=0
\ ,\ \ \
W,Y\ll 1
\,. 
\ee
The second relation allows to relate  the  future LHC bounds on  the HEP $a_{q}^{(3)}$ with the LEP bound  on the $\hat S$-parameter, providing an educated context to compare the impact of these two different machines.

There are also "Weak" theories that do not belong to the "universal" class, hence they must be described by a complete set of operators such as the Warsaw basis \cite{Grzadkowski:2010es}, see table~\ref{warsawoperators}.
 In this case, the HEP are transparently identified with contact interactions between quarks and scalars \footnote{These relations, as well as those in  \eq{a2c}, are obtained by computing the diboson helicity amplitudes in the presence of the EFT operators, and matching with the parametrization in \eq{primaries0}. The matching depends on the conventions for the spinor wave functions and the polarization vectors. We fix the ambiguity by reporting in Appendix~\ref{appA} the SM amplitudes computed with the same conventions.}
\be
a_u=4\frac{c_R^u}{M^2}\ ,\  \ a_d=4\frac{c_R^d}{M^2}\ ,\ \  a^{(1)}_q=4\frac{c^{(1)}_L}{M^2}\ ,\ \  a^{(3)}_q=4\frac{c^{(3)}_L}{M^2}\,.
\ee
Representatives of such ``non-universal'' theories are models with a heavy ${\textrm{SU}}(2)_L$ triplet vector boson $W'^a$ ($a=1,2,3$), coupled to the left-handed fermions and to the Higgs
\be\label{hvt}
{\cal L}_{\rm int}= \frac{1}{2}W'^a_\mu \left[g_f\bar f_L\gamma^\mu \sigma^a f_L+i g_H H^\dagger  \sigma^a \lra {D^\mu} H\right]\,,
\ee
where $g_f$ is in general different for the different SM fermions. In this type of models, after integrating out the heavy $W'^a$ at tree level, one obtains
\be\label{Wpred}
a_q^{(3)}= -\frac{g_qg_H}{M^2}\ ,\ \ \ \
a_q^{(1)}= a_u = a_d = 0\ ,\ \ \ \
\ee 
where $M$ is the mass of $W'^a$ and $g_q$ denotes the coupling to the light generation quark doublets.  In addition, there are also  induced  4-fermion interactions $g^2_f(f_L\gamma^\mu \sigma^a f_L)^2$ that are constrained, for the case of  quarks, by LHC high-energy di-jet experiments~\cite{Domenech:2012ai,Bellazzini:2017bkb}. Moreover, a shift in the fermion coupling to the $Z$ boson is also generated, that for the quarks reads $\delta g^Z_{uL}/g=-\delta g^Z_{dL}/g=-g_qg_Hv^2/(8c_{\theta_W}M^2)$ and is constrained mostly by LEP1 \cite{Pomarol:2013zra}.  This model can also be studied as an example of universal theory, in which case quarks and leptons couplings are equal because they emerge from the kinetic mixing of the heavy vector triplet with the SU$(2)_L$ SM gauge field strength. With the parameter scaling $g_f={\rm{c}}_F g^2/g_*$ and $g_H={\rm{c}}_H g_*$, they provide a simplified phenomenological description of composite Higgs vector resonances \cite{Pappadopulo:2014qza}. We will use this setup in section~\ref{sec:results} in order to compare the indirect reach from the HEPs with the one from direct resonance searches.

\section{LHC Primaries Sensitivity}\label{sec:sens}

LHC run-$2$ and $3$, the HL-LHC and future colliders can probe the HEP parameters. In this section we first work out a rough estimate of the reach in the  channels $Wh$, $Zh$, $WW$ and $WZ$; the result of this estimate will lead us to focus on fully leptonic $WZ$ that emerges as a particularly promising and simple option.

\subsection{Diboson Channels Overview}\label{overview}

From table~\ref{Wilsons}  we see that several diboson processes will have to be measured in order to get access to all the $4$ HEP parameters,   therefore we should not restrict only  to the channel with better reach.  
Nevertheless, it is convenient, as a starting point of a more complete analysis (that however goes beyond the scope of the present paper), to imagine probing a BSM scenario that produces comparable effects in all the channels, such that a comparison of the reach becomes relevant. A benchmark scenario of this sort is obtained by turning on the HEP parameter $a_q^{(3)}$, which enters in all the diboson processes, as table \ref{Wilsons} shows. In what follows we will thus focus on $a_q^{(3)}$ and compute the $95\%$ CL reach that is obtained in the various channels by a $\chi^2$ test on the distribution of the vector boson transverse momentum~$p_{T,V}$.\footnote{$p_{T,V}$ is defined here as the transverse momentum of any of the two bosons, which are equal in the tree-level simulations we employ in this section. The definition we will adopt in the more realistic analysis of section~\ref{sec:lWZ}  is given in eq.~(\ref{ptv}).} Only statistical uncertainties are included, assuming the full luminosity ($3\,$ab$^{-1}$) of the HL-LHC. Signal cross-sections are computed at tree-level using {\sc{MadGraph5}} v2.5.5~\cite{Alwall:2014bq} (and NNPDF 2.3LO1~\cite{Ball:2012cx} parton distributions) in the $p_{T,V}$ bins reported in table~\ref{tab:signal_bkg_estimate}. Only the interference contribution to the signal, i.e. the term linear in $a_{q}^{(3)}$, is reported in the table for shortness. Obviously the complete cross-section is used to derive the limit. The estimate of the background in each of the four channels will be described later.

The signal model was implemented in {\sc{MadGraph5}} by turning on the operator $\op_{HW}$ (defined in table~\ref{silhoperators}) in the model \texttt{EWdim6} of Ref.~\cite{Degrande:2012wf}, with a coefficient  $c_{HW}=\,a_{q}^{(3)}M^2/g^2$ as dictated by \eq{a2c}. We could have also implemented it through another operator, for instance $\op_{L}^{(3)}$, obtaining essentially identical results, since we have shown in the previous section that the high-energy cross-section is only sensitive to the HEP parameters. Indeed, we have checked that the discrepancy in the signal cross-sections, if $\op_{L}^{(3)}$ is employed (with $c_{L}^{(3)}=\,a_{q}^{(3)}M^2/4$, see again \eq{a2c}), is below $10\%$ for $p_{T,V}> 200\gev$ and around $1\%$ if $p_{T,V}>400\gev$. The discrepancy is due to the fact that the operators are of course not equivalent at finite energy, consequently it scales like $m_W^2/E^2$.
\begin{table}
\centering
{\small
\begin{tabular}{@{\hspace{.05em}}c@{\hspace{.35em}}c||c|c|c|c@{\hspace{.05em}}}
& & \multicolumn{4}{c}{$p_{T,V} $ range (GeV) }\\\hline
\multicolumn{2}{c||}{Channel} & [200, 400] & [400, 600] & [600, 1000] & [1000, 2000]\\
\hline\hline
\rule{0pt}{1.25em} & $W^{\pm}_{L}h$ & $23300 + 42500\, a_q^{(3)}$ & $1950 + 9750\, a_q^{(3)}$ & $420 + 4680\, a_q^{(3)}$ & \\
\rule{0pt}{1.25em} $W^{\pm}h$ & $W_{L}h$ substr.~\cite{Butterworth:2015bya} & $2230 + 4070\, a_q^{(3)}$ & $368 + 1840\, a_q^{(3)}$ & $108 + 1200\, a_q^{(3)}$ & \\
\rule[-.75em]{0pt}{2.em} & background~\cite{Butterworth:2015bya} & $11400$ & $1720$ & $700$ &\\
\hline
\multirow{2}{*}{$Zh$} & $Z_{L}h$ & $3760 + 5330\, a_q^{(3)}$ & $294 + 1350\, a_q^{(3)}$ & $58 + 600\, a_q^{(3)}$ & \\
\rule[-.75em]{0pt}{2.em} & $Z_{L}h$ substr.~\cite{Butterworth:2015bya} & $600 + 850\, a_q^{(3)}$ & $84 + 390\, a_q^{(3)}$ & $17 + 178\, a_q^{(3)}$ & \\
\hline
\multirow{2}{*}{$W^{+}W^{-}$} & \rule{0pt}{1.25em} $W_{L}W_{L}$ & $5080 + 7450\, a_q^{(3)}$ & $380 + 1730\, a_q^{(3)}$ & $74 + 780\, a_q^{(3)}$ & $5.8 + 160\, a_q^{(3)}$\\
\rule[-.75em]{0pt}{2.em} & other helicities & $89500$ & $5500$ & $990$ & $69$\\
\hline
\multirow{2}{*}{$W^{\pm}Z$} & \rule{0pt}{1.25em}  $W_{L}Z_{L}$ & $2970 + 5050\, a_q^{(3)}$ & $226 + 1200\, a_q^{(3)}$ & $46 + 540\, a_q^{(3)}$ & $3.7 + 123\, a_q^{(3)}$\\
\rule[-.75em]{0pt}{2.em}  & other helicities & $10800$ & $600$ & $100$ & $6.0$
\end{tabular}
}
 \caption{{Expected events at the $14$~TeV LHC with integrated luminosity $3\,$ab$^{-1}$ for the various diboson channels. The rates take into account the branching fractions $h\to b\bar{b}$, $W\to\ell\nu$ and $Z\to \ell\bar{\ell}$ with $l=\mu,e$. The number of events  in $W^{\pm}h$ and $Zh$  is negligible in the last bin. The value of $a_{q}^{(3)}$ is expressed in $\textrm{TeV}^{-2}$.}}
\label{tab:signal_bkg_estimate}
\end{table}

In the $WW$ and $WZ$ channels we considered leptonically decaying vector bosons, based on the fact that it is more difficult to perform accurate measurements in hadronic final states. The $b\bar{b}$ decay mode is instead considered for the Higgs in the $Wh$ and $Zh$ channels (with the vector bosons still leptonic), because fully leptonic Higgs decays are too rare to be relevant. Decay branching ratios are included in the cross-sections reported in table~\ref{tab:signal_bkg_estimate}. For $WW$ and $ZW$ it is not far from realistic to assume that all the reducible backgrounds can be neglected, and the only background for $W_{L}W_{L}$ and $Z_{L}W_{L}$ production arises from the production of the other polarization states (in particular the transverse $TT$). We see in table~\ref{tab:signal_bkg_estimate} that this background is sizable, and particularly so for $WW$. Hence the reach on $a_q^{(3)}$ (see table~\ref{tab:est_bounds}) is significantly better in $WZ$ than $WW$ after the background is included. For $W_{L}h$ and $Z_{L}h$ instead the background from the other polarizations is negligible since transverse vector boson plus Higgs production is suppressed at high energy. Reducible backgrounds (e.g., from $V$+jet or $t\bar{t}$ processes) are on the contrary sizable. For these processes we assume that boosted Higgs reconstruction will be performed with jet substructure techniques and we apply to the signal the Higgs reconstruction efficiency obtained in Ref.~\cite{Butterworth:2015bya}, where a careful analysis of the $Wh$ channel was performed. This efficiency varies from $\sim15\%$ in the low-$p_{T,V}$ bin to $\sim 25\%$ in the last bin, hence it entails a considerable loss of rate and in turn of sensitivity.\footnote{Actually, in the case of $Wh$ the ``substr.'' line in table~\ref{tab:est_bounds} also includes the efficiency of the jet veto cut of Ref.~\cite{Butterworth:2015bya}. The latter efficiency is however marginally relevant as it ranges from $60$ to $80\%$.} The $Wh$ background estimate is also taken from Ref.~\cite{Butterworth:2015bya}. Its impact on the reach is considerable, as shown in table~\ref{tab:est_bounds}, meaning that a significant improvement of boosted Higgs reconstruction techniques would be needed in order to make this channel competitive. We are not aware of detailed analyses focused on the high-$p_{T,V}$ regime of the $Zh$ process, therefore we studied this channel in the unrealistic hypothesis of no background. The reach in $Zh$ is slightly worse than the one in $WZ$ even in the absence of background because of the small leptonic $Z$ branching ratio. The background will further worsen the situation similarly to what happens in $Wh$. The two channels $Wh$ and $Zh$ are expected to face similar challenges for background reduction. 

We see that the fully leptonic $WZ$ process is expected to have the best reach among the channels we considered. Compared with associated Higgs production processes, it does not suffer from the large background due to boosted Higgs mistag and from the potentially sizable systematic uncertainties that could emerge when dealing with hadronic final states. Compared with $WW$, $WZ$ has a smaller background from transverse polarizations. This properties follows from a reduction of the transverse amplitude in the central region, as we will now discuss. While in what follows we will focus on this channel, it should be kept in mind that $WZ$ is only sensitive (see table~\ref{Wilsons}) to $a_q^{(3)}$, so that other channels will have necessarily to be studied in order to probe all the $4$ HEP parameters. We will further comment on this aspect in the Conclusions.

\begin{table}
\centering
\begin{tabular}{c|c c}
Channel & Bound without bkg. & \ Bound with bkg.\\
\hline
\rule{0pt}{1.25em} $Wh$ & $[-0.0096, 0.0096]$ & $[-0.036, 0.031]$\\
\rule{0pt}{1.25em} $Zh$ & $[-0.030, 0.028]$ & --\\
\rule{0pt}{1.25em} $WW$ & $[-0.012, 0.011]$ & $[-0.044, 0.037]$\\
\rule{0pt}{1.25em} $WZ$ & $[-0.013, 0.012]$ & $[-0.023, 0.021]$
\end{tabular}
 \caption{{Bounds on  $a_{q}^{(3)}$   (in $\textrm{TeV}^{-2}$) from the estimates of table~\ref{tab:signal_bkg_estimate}.
}}
\label{tab:est_bounds}
\end{table}

\subsection{Leptonic $\mathbf{WZ}$}\label{sec:lWZ}

The fully leptonic $WZ$ process 
$$
pp\to W^{\pm}Z + {\rm{jets}}\to \ell\nu\ell^{\prime}\bar{\ell}^{\prime} + {\rm{jets}} \,, \;\;\; {\rm{with}}\;\;l,l'=e,\mu\,,
$$ 
is likely to be measured with good accuracy. The leptons can be accurately reconstructed and the reducible background from other processes (which might hamper the whole procedure if not modeled well enough) is very low \cite{Aad:2016ett}. At the experimental level the situation might not be too much different from the neutral Drell-Yan process, in which a measurement with $2\%$ relative systematic uncertainty of the differential cross-section was performed, with run-$1$ data, up to TeV energies \cite{Aad:2016zzw}. A systematic uncertainty of $5\%$ might be considered as a realistic goal for the differential cross-section measurement in the leptonic $WZ$ channel. 

Since reducible backgrounds are under control, the main obstacle to obtain sensitivity to new physics is the potentially large contribution of the other polarizations, which for our purposes constitute a background, since they are insensitive to the new physics parameter $a_{q}^{(3)}$. In the $WZ$ channel these effects are automatically under control in the high-$p_T$ region and they can be further reduced by suitable selection criteria, as we will discuss later.

\subsubsection{Amplitude Zero}
In the SM, the longitudinally polarized final state $W_{L}Z_{L}$ is a subdominant fraction of the total cross-section. Indeed it accounts for just 6\% of the total rate for $pp\to WZ$ at the $14$~TeV LHC, which is dominated by transverse polarizations production. This is mainly due to the presence of a $t$-channel pole for the transverse polarizations (in particular the $+-$ and $-+$ helicity amplitudes) that significantly enhances the forward-scattering amplitude. Such contribution is absent for longitudinally polarized bosons. This forward enhancement is however tamed at high vector boson transverse momenta, where the amount of longitudinally polarized bosons becomes significantly larger, reaching a fraction $\sim 40\%$ of the total rate for $\ptv > 1\ {\rm TeV}$ (see table~\ref{tab:signal_bkg_estimate}).

A similar qualitative behavior is found in the $WW$ production process, however the high-$p_T$ cut is much less effective. We see in table~\ref{tab:signal_bkg_estimate} that in this case the longitudinal bosons are less than $10\%$ of the total for $\ptv > 1\ {\rm TeV}$. This is due to the fact that the high-energy $WZ$ amplitudes for the transverse $+-$ and $-+$ polarizations nearly vanish at tree-level if the bosons are produced centrally~\cite{Baur:1994ia}.\footnote{The production of same-sign diboson helicities, and of one transverse and one longitudinal boson, are anyhow suppressed at high energy, as we discussed in section~\ref{sec:thf}. Hence the suppression of the $+-$ and $-+$ amplitudes entails a suppression of the entire background 
cross-section.} The high-$p_T$ cut enhances the central region and consequently it reduces the transverse contribution more effectively in the $WZ$ channel than in the $WW$ one, where the central suppression of the transverse amplitudes is not present. Specifically, the $WZ$ tree-level amplitudes at high energy $E\gg m_W$ takes the form 
\begin{eqnarray} 
A(\bar u_L d_L \rightarrow W^-_{(\pm)} Z_{(\mp)}) \propto \cos \thetastar + \frac{1}{3} \tan^2 \theta_{\textsc w}\,,\nonumber\\
A(\bar d_L u_L \rightarrow W^+_{(\pm)} Z_{(\mp)}) \propto \cos \thetastar -\frac{1}{3} \tan^2 \theta_{\textsc w}\,.
\label{LOzwAmpTT}
\end{eqnarray}
where 
$\thetastar$ denotes the polar scattering angle in the collision rest-frame, oriented in the direction that goes from the incoming anti-quark to the outgoing $W$. Such behavior can be understood by symmetry arguments \cite{Frye:2015rba}. Since $\theta_{\textsc w}$ is small, the amplitude is suppressed at $\cos\thetastar \simeq 0$ (i.e. for central diboson production  $\thetastar \sim \pi/2$) and so  is the cross-section. Notice that this would not have been the case if the amplitude zero were not located at $\cos\thetastar \simeq 0$ because the $pp\to WZ$ differential cross-section $d\sigma/d\ptv$ is insensitive to the sign of $\cos \thetastar$, the two configurations with opposite $\cos \thetastar$ corresponding to the anti-quark coming from the first proton or from the second one. If the amplitude vanishes at $\cos \thetastar\neq0$, summing over the two configurations produces a differential cross-section that never vanishes. 

On the other hand, as expected from \eq{amp0}, the amplitude of the longitudinally polarized vector bosons is maximal at $\cos\thetastar\simeq 0$ because at high energy
\begin{equation}
A(u\bar d \rightarrow W^+_{L} Z_{L}) \propto \sin \thetastar\,, \label{LOzwAmp}
\end{equation}
both in the SM and when BSM effects are present. From eqs.~(\ref{LOzwAmpTT}) and (\ref{LOzwAmp}) we conclude that it could be advantageous to search for the effects of $\aq$ in the central region $|\cos\thetastar|<|\cos\thetastar|_{\rm{max}}$. Due to the fast decrease of the parton distribution functions at high energy, the region of phase-space where $\ptv$ is large tends to coincide with the region where $\cos \thetastar$ is small. Hence a centrality cut on $\cos \thetastar$ is already indirectly present in the large $\ptv$ bins as previously mentioned. Imposing it directly as $|\cos\thetastar|<|\cos\thetastar|_{\rm{max}}$, with $|\cos\thetastar|_{\rm{max}}$ to be determined, might still bring some improvement in the reach, as we will see.

Before describing in more details our selection criteria and their optimization (see section~\ref{selc}), we should however assess the robustness of our strategy with respect to NLO QCD correction. All the previous  considerations indeed rely on the amplitude zero, which is a tree-level effect that is lifted by QCD corrections. In section~\ref{rrc} we will investigate how the NLO real corrections affect the suppression of the production of transverse vector bosons in the central region. This will allow us to develop further insights for the design of our analysis strategy, which we apply in section~\ref{sec:nloanalysis} to the full NLO signal simulation.

\subsubsection{Real radiation corrections}\label{rrc}
In order to understand the structure of the NLO QCD corrections to $WZ$ production we first study real emissions. Namely, we consider the processes 
\begin{equation}
pp\to WZ\,, \;\;\;\;\; pp\to WZ\textrm{ + $1$ jet}\,, \label{eq:matchedprocess}
\end{equation}
simulated at tree-level and combined with QCD parton shower using the MLM scheme~\cite{Mangano:2006rw}.\footnote{Matrix elements for the calculation are computed with {\sc{MadGraph5}} and proton parton density functions  NNPDF 2.3LO1. The parton shower we used is Pythia6~\cite{Sjostrand:2006za} and jets are obtained from the shower results according to the $k_T$ clustering algorithm~\cite{Catani:1993hr}.} Real radiation is expected to be the most important correction to the amplitude zero described above, because extra parton emissions invalidate the symmetry arguments that one can make~\cite{Frye:2015rba} to explain the result in eq.~(\ref{LOzwAmpTT}).

\begin{figure}[t]
\centering
\includegraphics[width=0.475\textwidth]{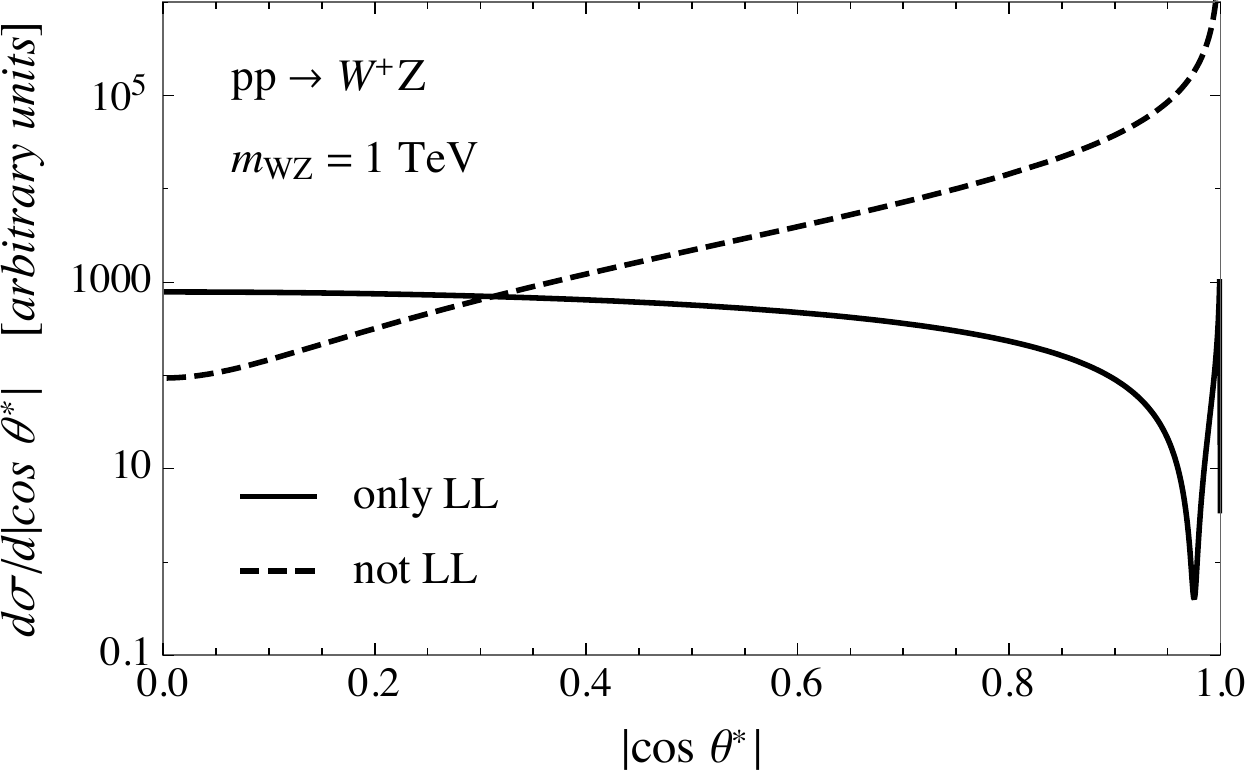}
\hfill
\includegraphics[width=0.475\textwidth]{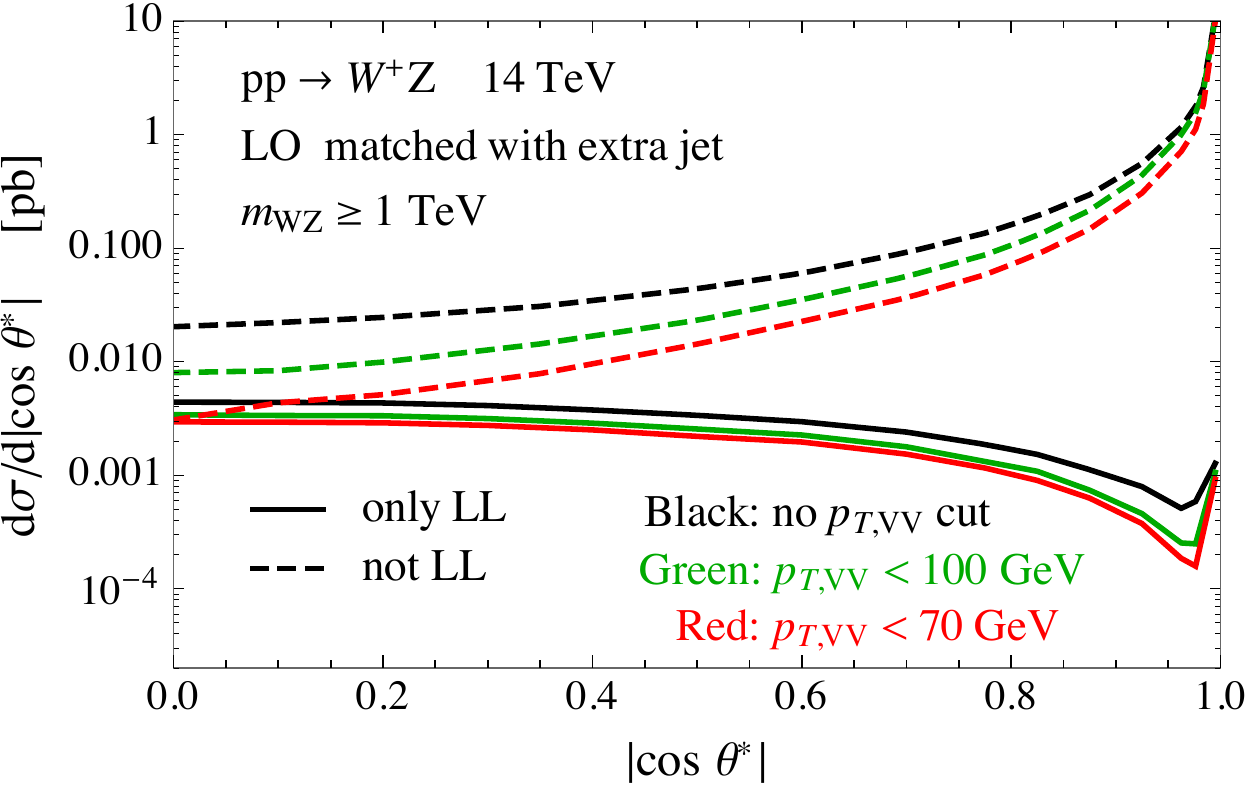}
\caption{Differential $\cos \thetastar$ cross-section for  $pp \rightarrow W^+ Z$, where  the solid (dashed) lines correspond to the final state with two longitudinally polarized  gauge bosons (all the other polarizations). Left: LO results at  invariant mass $m_{\textsc{wz}} = 1\ \rm{TeV}$. Right:  tree-level results  matched with an extra jet with  invariant mass  $m_{\textsc{wz}} \geq 1\ \rm{TeV}$.}
\label{Fig:WZ_LO_vs_matched}
\end{figure}

The effect of real radiation corrections on the $\cos\thetastar$ distribution can be gauged by looking at fig.~\ref{Fig:WZ_LO_vs_matched}. In the left panel of the figure we show the leading order distribution, with no extra jet, for $W_{L}Z_{L}$ production (solid line) separately from the sum of all the other polarization states (dashed line) at fixed center of mass energy $m_{\textsc{wz}} = 1\ \rm{TeV}$. The suppression of the transverse channels for $\cos \thetastar \simeq 0$ is clearly visible. For $\cos\thetastar=0$ the longitudinal channel cross-section is nearly one order of magnitude larger than the other channels. Real radiation is included in the right panel. We see that once real radiation is taken into account, the $W_{L}Z_{L}$ final state is much less prominent in the region at small $\cos\thetastar$.  Indeed, it is subdominant with respect to the total cross-section even at small $\cos\thetastar$ if no extra cut is performed (corresponding to the black  lines in the figure). 

In order to reduce the effects of hard real radiation, we employ a selection on the transverse momentum of the $WZ$ system, denoted by $$\deltapt=|{\vec{p}}_{T,W}+{\vec{p}}_{T,Z}|.$$ Alternatively, we might have considered a jet veto, which however would have been problematic for accuracy because of the experimental and theoretical uncertainties associated with jets reconstruction. The $\deltapt$ variable is instead inclusive over the hadronic final state and it does not require jet reconstruction.\footnote{See Ref.~\cite{Campanario:2014lza} for a different approach.} The impact of the $\deltapt$ cut on the $\cos\thetastar$ distributions is displayed in the right panel of fig.~\ref{Fig:WZ_LO_vs_matched} for $\deltapt < 100\ \rm{GeV}$ (green) and $\deltapt < 70\ \rm{GeV}$ (red). We see that a requirement on $\deltapt$ significantly enhances  $pp\to Z_{L}W_{L}$ with respect to the background at low $\cos\thetastar$, but it does not make the background negligible. Notice that $\deltapt$ being an inclusive quantity does not necessarily mean that its distribution will be accurately described by a fixed-order QCD calculation. In particular if $\deltapt$ is much smaller than the bosons momenta, corresponding to a configuration where real soft radiation is nearly absent, one would need to perform resummation, which might bring additional uncertainties. We will take this potential issue into account when discussing the $\deltapt$ cut optimization in the following section.

\subsubsection{Optimization of the Selection Criteria}\label{selc}
Our strategy is to probe $a_q^{(3)}$ by performing a fit of the $p_{T,V}$ differential cross-section, where $p_{T,V}$ is defined as
\be
p_{T,V} = \min(p_{T,W}, p_{T,Z})\,.\label{ptv}
\ee 
Using the minimum momentum suppresses, in the large $p_{T,V}$ bins, configurations where one of the bosons is much harder than the other one and recoils against a jet. Since hard real radiation suppresses the signal relative to the background, those configurations are not relevant for our analysis. We will assume that the measurement of the $p_{T,V}$ cross-section will be performed in a fiducial region defined by selections cuts on $|\cos \theta^*|$ and on $\deltapt$. Selection criteria on these variables are indeed expected to improve the sensitivity as previously explained.\footnote{Measuring $|\cos \theta^*|$ requires neutrino reconstruction and introduces an ambiguity. We momentarily assume perfect neutrino reconstruction, postponing to section~\ref{sec:nloanalysis} the discussion of this point.} We now want to study more quantitatively the impact of these cuts, optimizing them in order to maximize the reach on $a_q^{(3)}$. For this purpose we consider three $p_{T,V}$ bins
$$p_{T,V} \in\{200, 400, 600,1000\}\ {\rm GeV}\,,$$
and we estimate the sensitivity to $a_q^{(3)}$ in each bin. We assume a $5\%$ systematic error in each bin, which we regard as a plausible goal for these measurements, whereas we neglect reconstruction efficiencies. We employ the LO matched simulation described in the previous section to compute the number of longitudinally-polarized events in each bin ($N_{LL}$) and the total ($N_{TOT}$) expected in the SM. The full HL-LHC luminosity is assumed. We estimate as $N_{LL}/\sqrt{N_{TOT} + (5\%\, N_{TOT})^2}$ the relative accuracy on the measurement of the longitudinally-polarized component of the cross-section in each bin. Since the effect of $a_q^{(3)}$ on the longitudinal cross-section grows quadratically with the energy, the relevant quantity to be computed in order to compare the sensitivity to $a_q^{(3)}$ of the different bins is not the accuracy of the measurement, but the accuracy rescaled by $(1/\ptvmin)^2$, where $\ptvmin$ is the lower endpoint of the bin. 

\begin{figure}[t]
\centering
\includegraphics[width=0.48\textwidth]{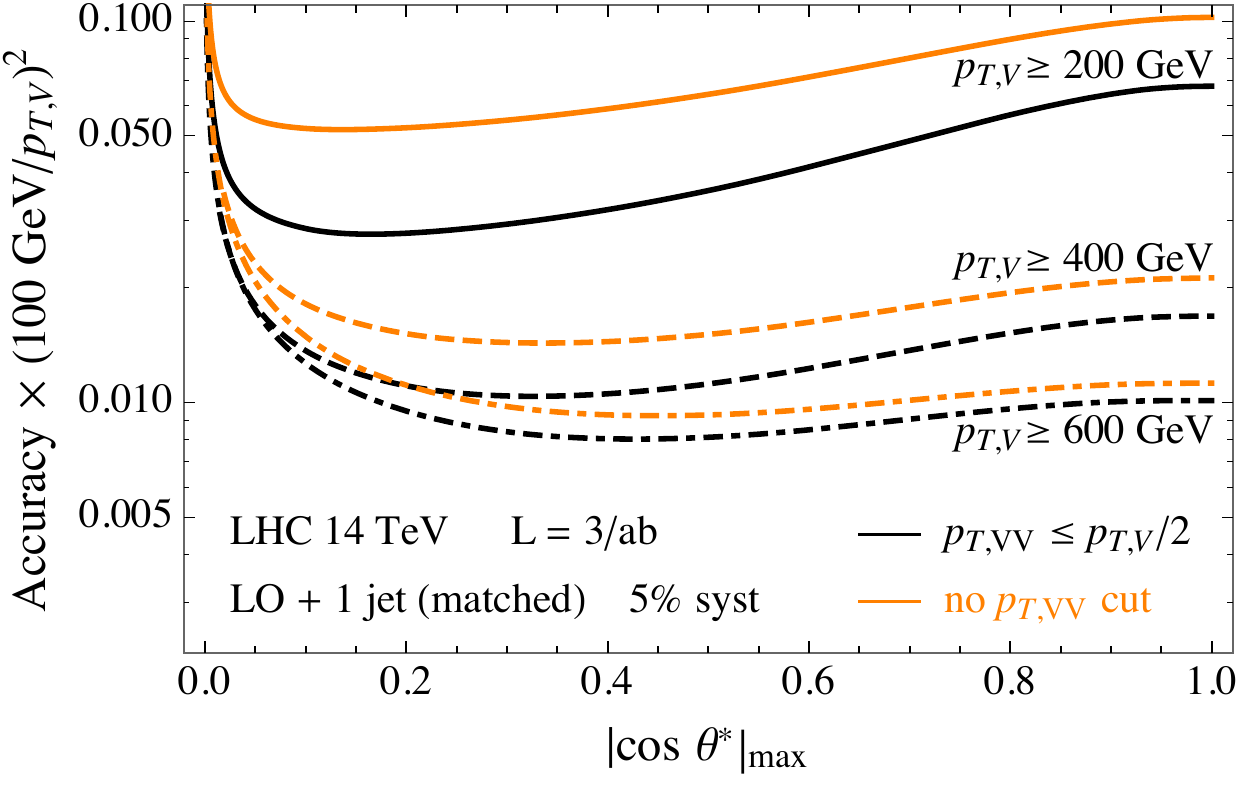}
\hfill
\includegraphics[width=0.47\textwidth]{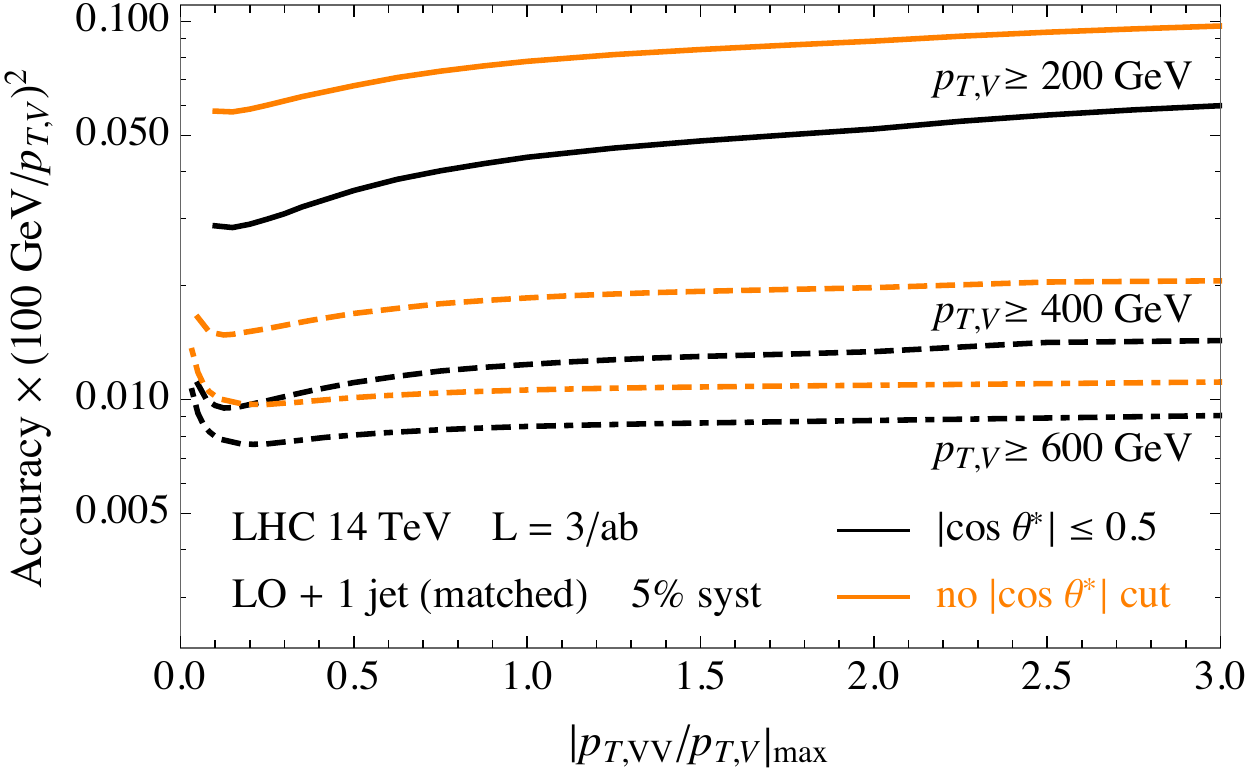}
\caption{Rescaled accuracy as a function of the cut on the scattering angle $|\cos \thetastar|$ (left panel) and of the transverse momentum of the $WZ$ system
$\deltapt$ (right panel). The solid, dashed, dot-dashed and dotted lines correspond to the three $p_{T,V}$ bins described in the main text.
The black lines are obtained by imposing the additional cuts $p_{T,VV} \leq p_{T,V}/2$ in the left plot and $|\cos \theta^*| < 0.5$
in the right plot. The orange lines are obtained with no additional cut.}
\label{Fig:WZ_cuts}
\end{figure}

The left panel of fig.~\ref{Fig:WZ_cuts} displays the rescaled accuracy as a function of the upper cut on $|\cos\thetastar|$, denoted as $|\cos\thetastar|_{\rm{max}}$. The curve has a mild dependence on the cut, aside from the low $|\cos\thetastar|_{\rm{max}}$ region where the lack of statistics reduces the sensitivity. The dependence of the rescaled accuracy on $|\cos\thetastar|_{\rm{max}}$ is also very mildly sensitive to the $\deltapt$ cut; for definiteness we use $\deltapt \leq p_{T,V}/2$ in the figure (black lines) and we include for comparison the results without $\deltapt$ cut (orange lines). For simplicity in what follows we use the cut
\begin{equation}
\label{costhetacut}
|\cos \thetastar| \leq|\cos \thetastar|_{\rm{max}}= 0.5\,,
\end{equation}
independently of $p_{T,V}$. We see that indeed this choice nearly minimizes the rescaled accuracy (hence it maximizes the reach) in the highest $p_{T,V}$ bin where the sensitivity is better, and is not far from the optimal choice for the $p_{T,V} \geq 400\;$GeV bin.
A harder cut would be required to maximize the sensitivity in the lowest bin, however the accuracy in this bin is quite poor,
so our simple choice of a $p_{T,V}$-independent cut does not significantly affect the reach.

As far as $\deltapt$ is concerned, we instead employ a $p_{T,V}$-dependent cut, namely 
\begin{equation} \label{ptvetoratio}
\deltapt/p_{T,V}<[\deltapt/p_{T,V}]_{\rm{max}}= 0.5\,.
\end{equation}
The dependence of the rescaled accuracy on $[\deltapt/p_{T,V}]_{\rm{max}}$ is very mild, as the right panel of fig.~\ref{Fig:WZ_cuts} shows. The chosen value of $0.5$ is slightly above the absolute minimum for the relevant $p_{T,V}$ bins, however this does not entail a significant loss of sensitivity. We took it somewhat larger than the minimum because it could be difficult to obtain accurate predictions for a too  low $\deltapt$ cut, as previously explained. Choosing $[\deltapt/p_{T,V}]_{\rm{max}}=0.5$ should leave enough phase space to real emission and allow for trustable fixed-order QCD calculations. Indeed we will verify in section~\ref{sec:nloanalysis} that scale uncertainties are not enhanced by this cut, while they would increase significantly if a tighter selection was adopted.

There are a few additional insights that can be extracted from the plots in fig.~\ref{Fig:WZ_cuts}. First of all it can be seen that the bins with $p_{T,V} > 400\ {\rm GeV}$ and $p_{T,V} > 600\ {\rm GeV}$ have the best, and comparable, sensitivity. They are followed by the bin $p_{T,V} > 200\ {\rm GeV}$, whose sensitivity is roughly a factor $4$ lower. This means that a possible new physics effect in this channel would not show up as a departure from the SM prediction which is localized in a single bin, but rather as a (arguably more convincing) tension with the SM distributed over a wide energy range. Second, from the figure we see that the cuts we devised increase the accuracy of around $30\%$ in the highest bin, $50\%$ in the intermediate and $70\%$ in the lowest. We checked that this is mainly due to the reduction of the signal over background ratio that mitigates the impact of systematic uncertainties.

\subsubsection{NLO Analysis \label{sec:nloanalysis}}


\begin{table}[t]
\begin{centering}
\begin{tabular}{c|c}
\rule[-.5em]{0pt}{1.em}$p_{T,V}$ range & Expected Events 
\tabularnewline
\hline 
\hline 
\rule[-.5em]{0pt}{1.6em}{[}100-150{]} GeV & 3100 + 1040 $a_{q}^{(3)}$ + 260 ${a_{q}^{(3)\;2}}$ 
\tabularnewline
\hline 
\rule[-.5em]{0pt}{1.6em}{[}150-220{]} GeV & 2620 + 1030 $a_{q}^{(3)}$ + 140 ${a_{q}^{(3)\;2}}$ 
\tabularnewline
\hline 
\rule[-.5em]{0pt}{1.6em}{[}220-300{]} GeV & 937 + 600 $a_{q}^{(3)}$+ 230 ${a_{q}^{(3)\;2}}$ 
\tabularnewline
\hline 
\rule[-.5em]{0pt}{1.6em}{[}300-500{]} GeV & 544 + 700 $a_{q}^{(3)}$ + 560 ${a_{q}^{(3)\;2}}$ 
\tabularnewline
\hline 
\rule[-.5em]{0pt}{1.6em}{[}500-750{]} GeV & 86.5 + 260 $a_{q}^{(3)}$ + 490 ${a_{q}^{(3)\;2}}$ 
\tabularnewline
\hline 
\rule[-.5em]{0pt}{1.6em}{[}750-1200{]} GeV & 16.1 + 120 $a_{q}^{(3)}$ + 640 ${a_{q}^{(3)\;2}}$ 
\end{tabular}\par\end{centering}
 \caption{{\label{NevCW}Expected number of events as a function of the HEP $a_{q}^{(3)}$ (expressed in $\textrm{TeV}^{-2}$) in each bin of the $p_{T,V}$ spectrum at LHC 14 TeV for 3/ab integrated luminosity.}}
\end{table}

We now estimate the reach on $a_q^{(3)}$ based on a full NLO simulation of the $pp \to 3\ell\nu$ process. We perform a matched calculation that uses matrix elements computed at NLO in QCD with {\sc{MadGraph5}} with FxFx-matched \cite{Frederix:2012ly} parton shower supplied by Pythia8~\cite{Sjostrand:2014rr}, with NNPDF~2.3~NLO parton distributions. The signal is computed  (as explained in section~\ref{overview}) through the operator $\op_{HW}$ implemented in the NLO version of the UFO model \texttt{EWdim6}, kindly provided to us by C.~Degrande. We consider generation-level leptons momenta, but we include an overall detector efficiency for reconstructing the three leptons that, based on performances studies in Refs.~\cite{ATLAS-CONF-2016-024,ATLAS-Collaboration:2016aa}, we estimate around 50\%. We furthermore apply standard acceptance cuts
\be 
p_{T,\ell}>30 \gev \ ,\quad |\eta_{\ell}|<2.4\,. \label{leptonscuts}
\ee 
The same-flavor and opposite-charge lepton pair with invariant mass closer to the $Z$ boson mass is taken as the $Z$ candidate and the remaining lepton  is taken to be the decay product of the $W$ boson. The missing transverse energy vector of the event ($\cancel{\vec{E}}_T$ ) is estimated from the generation-level $x$ and $y$ neutrino momentum components, to which we apply a Gaussian smearing with standard deviation
$$ \sigma_{{\met}_{i}}^2 = (0.5)^2 \cdot \sum_f |p_i| \cdot {\rm{GeV}}\,.$$ 
This approach is similar to well-tested detector performance parameterizations used e.g. in {\sc{Delphes}}~\cite{Ovyn:2009ys,de-Favereau:2013dk}.

The kinematical variables described so far allow us to determine $p_{T,Z}$ and $p_{T,W}$, and in turn $p_{T,V}$ and $p_{T,VV}$, used to construct the binned distribution and for the selection cut in eq.~(\ref{ptvetoratio}), respectively. In order to extract $|\cos \thetastar|$, which we will employ for the selection in eq.~(\ref{costhetacut}), the reconstruction of the neutrino rapidity is needed. This is obtained by the standard technique of imposing the invariant mass of the neutrino plus lepton system to be as close as possible to the physical $W$ boson mass. If the lepton transverse mass $m_{T\ell\nu}$ is smaller than $m_W$, the lepton-neutrino invariant mass can be asked to be equal to $m_W$, producing two solutions
\begin{equation}\label{eq:nu_reco_2}
\eta_\nu^{\pm} =\eta_\ell  \pm \log \left(1 + \Delta + \sqrt{\Delta(\Delta + 2)}\right)\,,\;{\rm{where}}\;\; \Delta\equiv\frac{m_{W}^2 - m_{T\ell\nu}^2}{2 p_T^\ell \met} \,.
\end{equation}
If instead $m_{T\ell\nu}>m_W$, which might happen because of experimental uncertainties in the measurement of the $\cancel{\vec{E}}_T$, or because the virtual $W$ had truly an invariant mass slightly above $m_W$, the lepton-neutrino invariant mass cannot be equal to $m_W$. The configuration that makes it as close as possible to $m_W$ is
\begin{equation}\label{eq:nu_reco_1}
\eta_\nu = \eta_\ell\,.
\end{equation}
If the $W$ is boosted in the transverse plane, which is the case in the kinematical region that is mostly relevant for our analysis, the reconstructed neutrino momentum becomes close to the true one both in the one-solution and in the two-solutions cases (see for instance \cite{Panico:2017??} for a recent discussion). However in the latter case we still formally have a twofold ambiguity in the determination of $\eta_\nu$, which in turn produces an ambiguity in $|\cos \thetastar|$. We resolve this ambiguity by imposing the cut in eq.~(\ref{costhetacut}) on both solutions, i.e. by retaining for the analysis only events such that both the possible neutrino configurations satisfy the selection criteria.

We study the $3$ collider energy options that correspond to the LHC ($14$~TeV), to the High-Energy LHC (HE-LHC, $27$~TeV) and to the FCC-hh ($100$~TeV). In each case we consider suitably designed $p_{T,V}$ bins, namely 
\begin{eqnarray}\label{finalbins}
{\textrm{LHC\hspace{-10pt}}}&{\textrm{:}}& p_{T,V}\in\{100,150,220,300,500,750,1200 \}\,,\\
{\textrm{HE-LHC\hspace{-10pt}}}&{\textrm{:}}& p_{T,V}\in\{150,220,300,500,750,1200, 1800 \}\,,\nonumber\\
{\textrm{FCC\hspace{-10pt}}}&{\textrm{:}}& p_{T,V}\in\{220,300,500,750,1200, 1800, 2400 \}\,.\nonumber
\end{eqnarray}
The binning is chosen such as to cover the kinematical regime that is accessible at each collider and it is taken as fine as possible in order to maximize the BSM sensitivity. On the other hand, a minimum bins size $\Delta p_{T,V}/ p_{T,V}\gtrsim30\%$ is required in order to avoid a degradation of the accuracy due to the $p_{T,V}$ resolution. After applying the selection cuts previously described, we compute the cross-section in each of the above bins and we fit it to a quadratic function of $a^{(3)}_q$. The results, expressed in terms of expected bin counts for $\mathcal{L}=3\,\textrm{ab}^{-1}$, are reported in table~\ref{NevCW} for the illustrative case of the $14$~TeV LHC. 

The predicted cross-sections are used to construct the $\chi^2$, under the assumption that observations  agree with the SM, and are eventually used to derive $95\%$~CL bounds on $a_q^{(3)}$. The uncertainties in each bin are the sum in quadrature of the statistical error, obtained from the SM expected events yield, and of a systematical component (uncorrelated across bins) which we take as a fixed fraction ($\delta_{\textrm{syst}}$) of the SM expectations. With this procedure we obtain, for different collider energies and luminosities and for $\delta_{\textrm{syst}}=5\%$
\bea
{\textrm{LHC, }} 300\,{\textrm{fb}}^{-1}\hspace{-10pt} & {\textrm{:}} & a^{(3)}_q  \in  [-1.4, 0.9] \,10^{-1}\,\TeV^{-2}\nonumber\\
{\textrm{HL-LHC, }} 3\,{\textrm{ab}}^{-1}\hspace{-10pt} & {\textrm{:}} & a^{(3)}_q  \in  [-4.9, 3.9] \,10^{-2}\,\TeV^{-2}\nonumber\\
{\textrm{HE-LHC, }} 10\,{\textrm{ab}}^{-1}\hspace{-10pt} & {\textrm{:}} & a^{(3)}_q  \in  [-1.6, 1.3] \,10^{-2}\,\TeV^{-2}\nonumber\\
{\textrm{FCC-hh, }} 20\,{\textrm{ab}}^{-1}\hspace{-10pt} & {\textrm{:}} & a^{(3)}_q  \in  [-7.3, 5.7] \,10^{-3}\,\TeV^{-2}
\label{boundsf}
 \eea 
We see that the HL-LHC will improve the LHC reach by more than a factor of $2$, while with the HE-LHC one would gain nearly one order of magnitude. A gain of around $20$ would be possible with the FCC-hh collider. The FCC-hh reach is comparable with the one of CLIC, as extracted from the analysis in Ref.~\cite{Ellis:2017kfi}.
 
Notice that the choice $\delta_{\textrm{syst}}=5\%$ is not based on a careful assessment of the experimental systematical uncertainties and of the theory errors on the SM predictions. At the experimental level, we merely argued at the beginning of section~\ref{sec:lWZ} that $\delta_{\textrm{syst}}=5\%$ could be a reasonable target, based on analogies with other purely leptonic final states. For what concerns theory, we verified that parton luminosity uncertainties are well below $5\%$ in the energy range of interest and that the scale variations in the NLO calculation are of order $5\%$. Scale variations were estimated using MCFM~8.0~\cite{Boughezal:2016ek,Campbell:2011fp} by varying renormalization and factorization scale as $\mu_{R}=\mu_{F}=2^{\pm 1} (m_{W}+m_{Z})$. A kinematic-dependent choice of the scales, e.g. $\mu_{R}=\mu_{F}=2^{\pm1}m_{\textsc{wz}}$ gives similar results.\footnote{We also checked that a tighter $\deltapt/p_{T,V}$ cut, such as $[\deltapt/p_{T,V}]_{\rm{max}}= 0.1$, would inflate scale uncertainties to the $20\%$ level. This had to be expected, as discussed in sections~\ref{rrc} and \ref{selc}.} Taking also into account that QCD NNLO \cite{Campanario:2010xn,Grazzini:2017ckn} and EW NLO \cite{Accomando:2005xp} computations are already available, we conclude that $\delta_{\textrm{syst}}=5\%$ or less is a reasonable target for theory uncertainties as well. We will discuss later in this section how a larger or a smaller value of $\delta_{\textrm{syst}}$ would affect the reach.
 
\begin{figure}[t]
\centering
\includegraphics[width=0.7\textwidth]{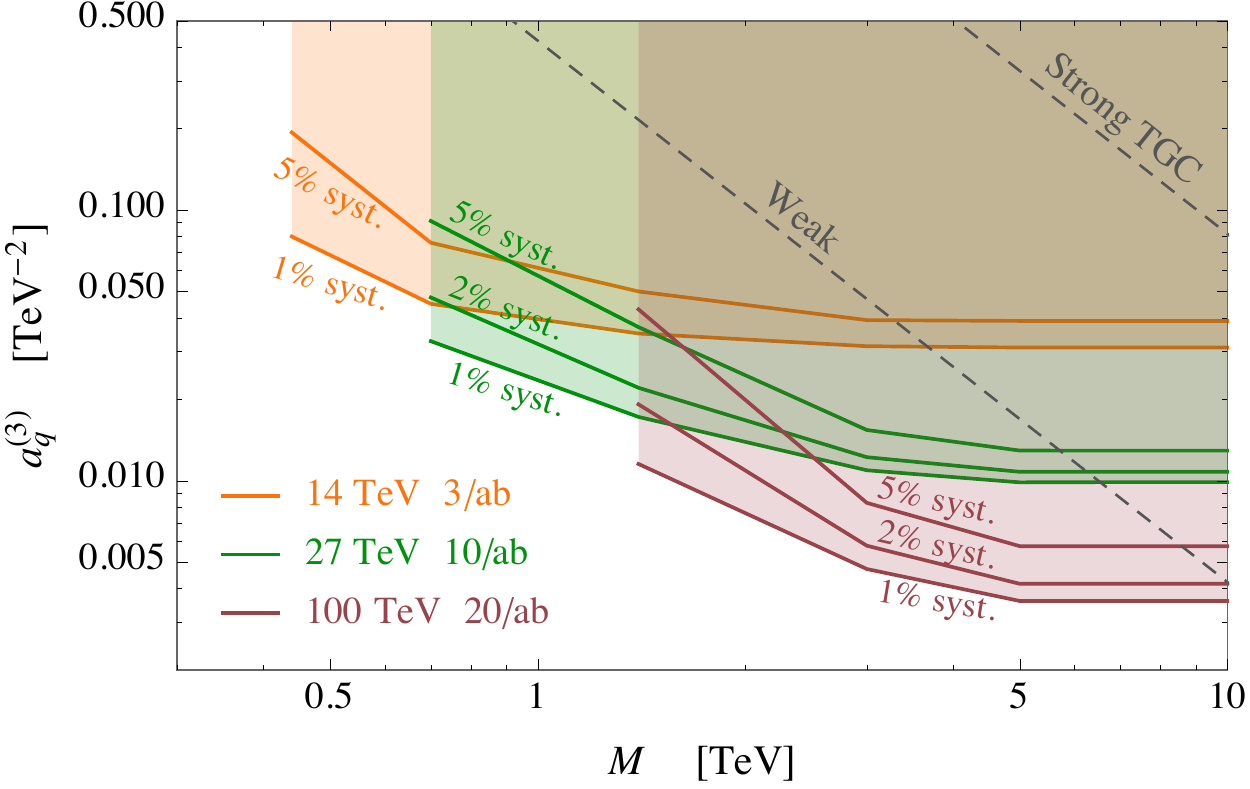}
\caption{Expected $95\%$ CL bounds from fully leptonic $WZ$ on the high-energy primary parameter $a^{(3)}_q$ as a function of the new physics scale $M$. The plots reports the results for the HL-LHC (orange lines), HE-LHC (green lines) and FCC-hh (brown lines) for different
values of the systematic uncertainties.} 
\label{fig:bounds_future}
\end{figure}

The results of \eq{boundsf}  rely on BSM cross-section predictions obtained by integrating up to very high center of mass energies, formally up to the collider threshold. Therefore these limits assume that the description of the underlying BSM model offered by the EFT is trustable in the whole relevant kinematical regime, i.e. that the cutoff $M$ of the BSM EFT is high enough. In other words, we assume that other effects such as resonance production, not included in the EFT description, take place to such a large $M$ that are irrelevant. We quantify how large $M$ concretely needs to be for our results to hold by studying \cite{Racco:2015dxa,Pobbe:2017wrj,Biekoetter:2014jwa} how the limit deteriorates if only events with low $WZ$ invariant mass, $m_{\textsc{wz}}<m_{\textsc{wz}}^{\rm{max}}$ are employed. This obviously ensures that the limit is consistently set within the range of validity of the EFT provided the EFT cutoff $M$ is below $m_{\textsc{wz}}^{\rm{max}}$.\footnote{The choice of the kinematical variable that best characterizes the hardness of the event, to be compared with $M$ in order to ensure the EFT validity, is ambiguous to some extent. One choice could be the total invariant mass of all the final state hard objects~\cite{Pobbe:2017wrj}, which in our case would include extra hard jets. The diboson mass $m_{\textsc{wz}}$ that we employ here is also a reasonable choice, in light of the cut on $p_{T,VV}$ that effectively vetoes hard QCD radiation.} The results are reported in figure~\ref{cvsm} for the LHC and the HL-LHC and in figure~\ref{fig:bounds_future} for the higher energy future collider options. Since the $95\%$~CL interval is nearly symmetric around the origin (with the exception of the LHC one), only the upper limit is reported in the figure for shortness.

Several conclusions can be drawn from figures~\ref{cvsm} and \ref{fig:bounds_future}. First of all we see that the reach saturates for $m_{\textsc{wz}}^{\rm{max}}$ below around $1.5$~TeV at the LHC and at the HL-LHC if the systematic uncertainties are low, meaning that the limits obtained without $m_{\textsc{wz}}$ cut apply to theories with cutoff $M$ above that threshold.  The threshold grows to around $3$ and $4$~TeV at the HE-LHC and at the FCC-hh, respectively. The curve with $\delta_{\textrm{syst}}=100\%$ in figure~\ref{cvsm} outlines the crucial role played by accuracy in this analysis.  An inaccurate determination of the cross-section would not only weaken by a factor $\sim4$ the asymptotic reach at  $m_{\textsc{wz}}^{\rm{max}}\rightarrow\infty$, but it would also raise above $2$~TeV the energy scale that is relevant for the limit. This makes that on one hand we would be only sensitive to theories with a lower $M$, since $a_q^{(3)}\sim1/M^2$, while on the other hand we would need theories with larger $M$ for our limit to hold. The combination of these two effects would drastically reduce the set of BSM theories that we would be able to probe.  This is illustrated in the figures by overlying to the reach the theoretical estimates of $a^{(3)}_q$, as a function of $M\simeq m_{\textsc{wz}}^{\rm{max}}$, in the ``Fully Strong'', ``Strong TGC and ``Weak'' scenarios described in the introduction and in section~\ref{bsmsection}. The fact that the $\delta_{\textrm{syst}}=100\%$ limit lies above or on top of the ``Weak'' line means that with this large systematic we can probe a given value of $a^{(3)}_q$ only if we trust the EFT prediction  at or above the cutoff of the ``Weak'' BSM theory that is producing that value, which is clearly inconsistent.  If instead $\delta_{\textrm{syst}}$ is low the reach stays well below the ``Weak'' line, meaning that we can probe BSM theories of the ``Weak'' type by only using events with a center of mass energy that is below the cutoff, for which the EFT description applies. The figures show that $\delta_{\textrm{syst}}=5\%$ is sufficient to probe ``Weak'' theories in all cases, but it also shows that the impact of a larger or smaller uncertainties on the reach is different at different colliders. In particular we see that the reach is very stable with $\delta_{\textrm{syst}}$ at the LHC, given that the $\delta_{\textrm{syst}}=10\%$ curve is very close to the one at $\delta_{\textrm{syst}}=1\%$, while it is much less so at the HL-LHC, where $\delta_{\textrm{syst}}=5\%$ already makes an appreciable difference with respect to $\delta_{\textrm{syst}}=1\%$. This is due to the fact that the low-$p_{T,V}$ bins are more populated at the HL-LHC, hence the statistical error is lower and the reach in those bins benefits from a lower systematics. The effect is even more pronounced at the HE-LHC and at the FCC-hh, where even with $\delta_{\textrm{syst}}=2\%$ the reach deteriorates significantly with respect the ideal case $\delta_{\textrm{syst}}=1\%$. The fact that more accurate measurements would improve the reach of future colliders is an element that should be taken into account in the design of the corresponding detectors.


\section{BSM Implications\label{sec:results}}

The impact of our results can be appreciated by  direct comparison of our bounds with bounds from other experiments. While HEP effects are uniquely probed at the LHC, we have seen in section~\ref{bsmsection} that, at the level of the dimension-six Lagrangian, HEP parameters are related to other parameters that can be measured in low-energy experiments. For a more specific BSM assumptions, we can have  more relations among observables.
We can for instance consider  a  BSM that affects mainly the  $Z$ couplings to quarks ${\delta g^Z_{q}}$ (for example, a theory with extra vector-like quarks that mix with the SM ones); in this case, we can read from table~\ref{Wilsons} that our constraint on $a^{(3)}_q$ corresponds to an impressive per-mille constraint on ${\delta g^Z_{q}}$. Such precision is competitive with LEP, that tested  ${\delta g^Z_{q}}$ by measuring precisely the $e^+e^-\to \bar q q$ differential cross-section on the $Z$ resonance.

\begin{figure}[t]
\centering
\includegraphics[width=0.48\textwidth]{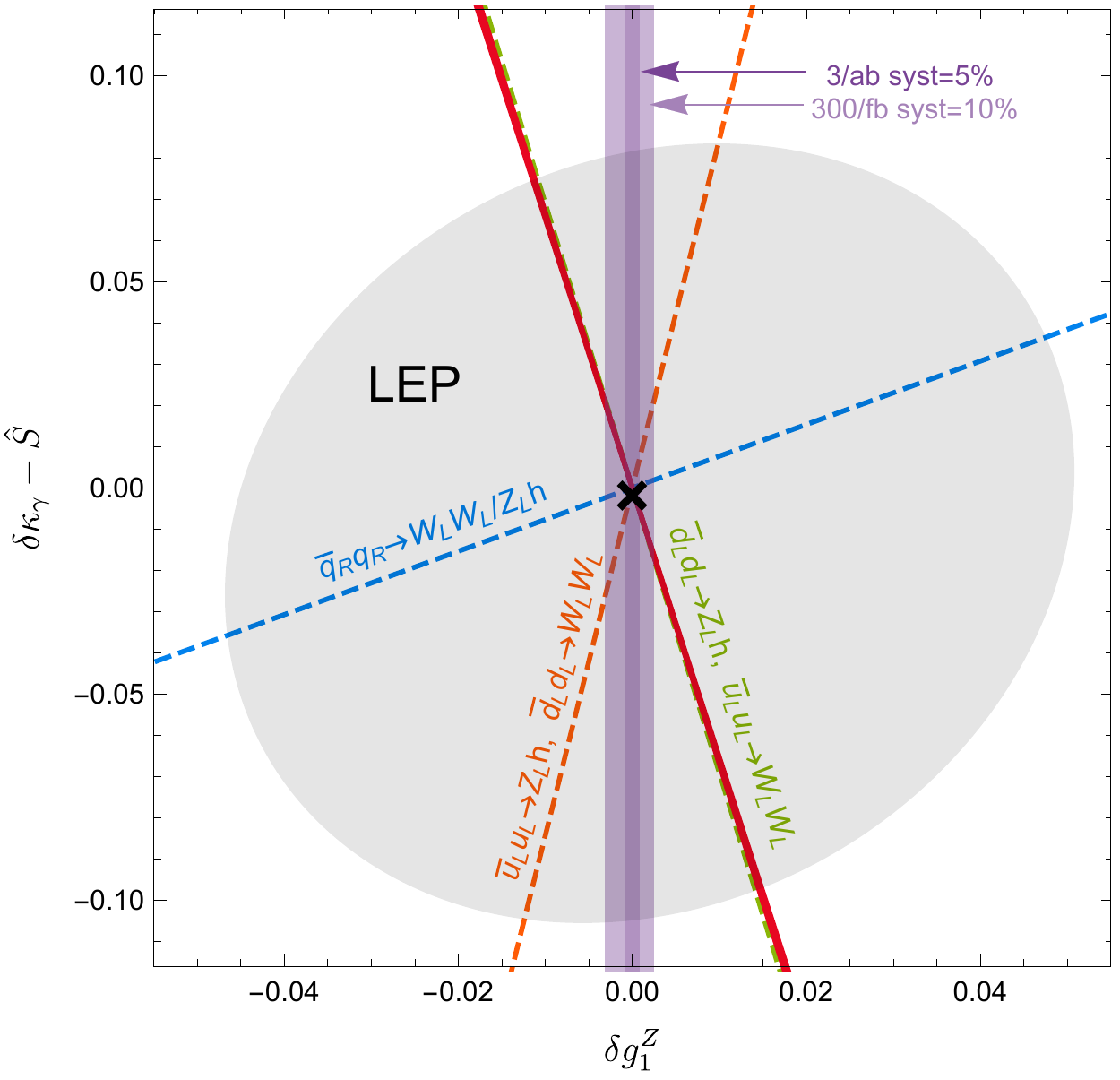}\hspace{0.5cm}
\includegraphics[width=0.48\textwidth]{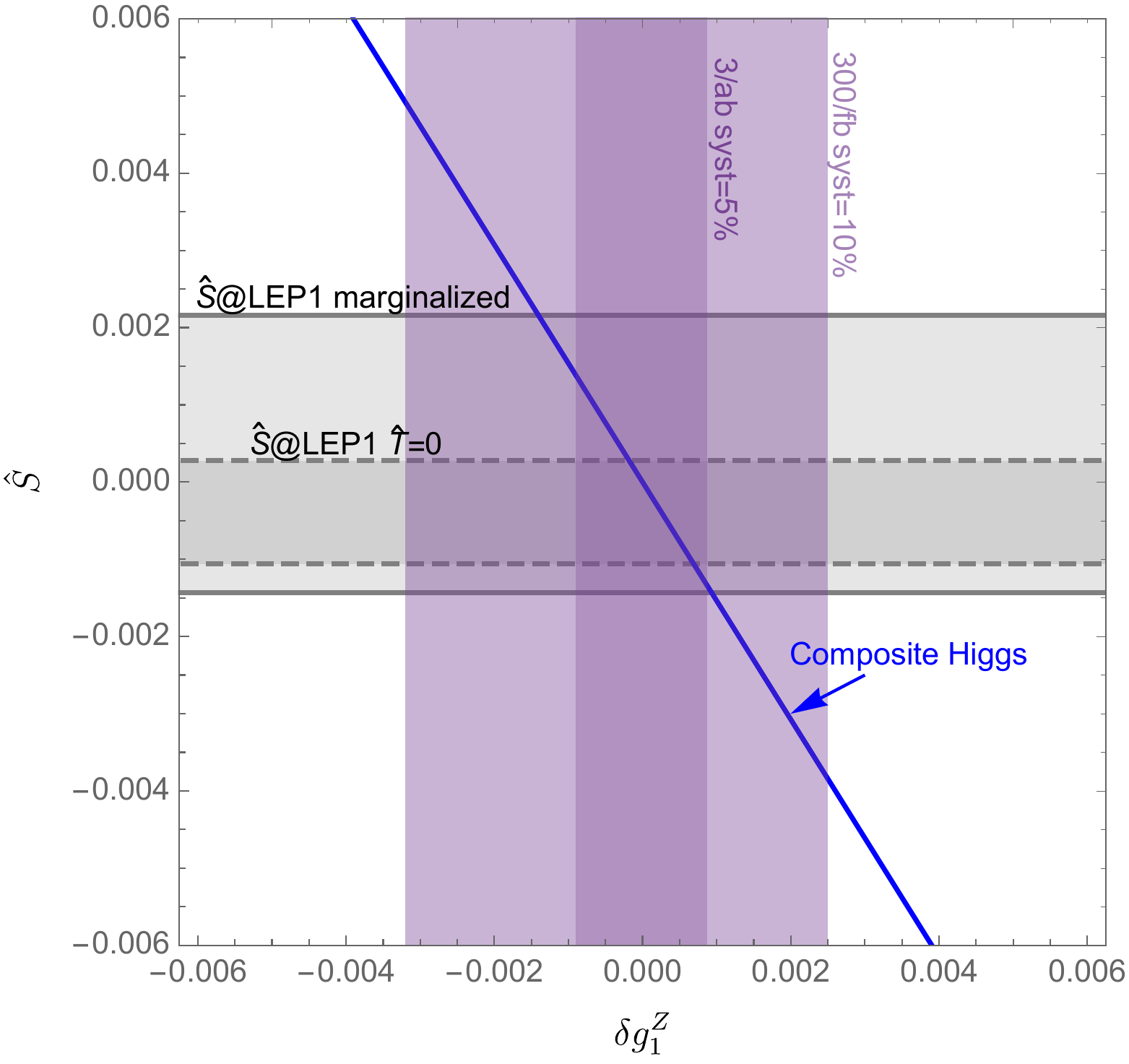}
\caption{Comparison of the   bounds obtained from LEP
with those from our analysis based on the $WZ$ channel at the LHC. 
Left:   universal theories with   $W,Y\ll 1$.
Right: Theories characterized by $W,Y,\delta \kappa_\gamma,\lambda_\gamma\ll 1$. See main text for details.} 
\label{cvsm2}
\end{figure}

Universal theories provide  an interesting framework to perform our comparison. There, \eq{heptosilh}  shows that indeed HEP parameters can be related to  $\hat S$, $W$, $Y$ \cite{Barbieri:2004qk} and the aTGCs $\delta g_1^Z$, $\delta\kappa_\gamma$. 
Now, $W$ and $Y$ characterize $O(p^4)$ corrections to the $W^\pm$ and $Z$ propagators, and can be splendidly tested  (well below the per-mille) at the HL-LHC , by  measurements of the dilepton invariant mass spectrum in charged and neutral Drell-Yan processes~\cite{Farina:2016rws}. In light of this, we can neglect  the effect of $W$ and $Y$ in our analysis. Then, from  \eq{heptosilh}, we see that HEPs overlap only with the $\delta g_1^Z$ and $ \delta \kappa_\gamma-\hat S$ combinations. 
In this two-dimensional parameter space, the $WZ$ channel, that we have studied in detail in section~\ref{sec:lWZ}, gives  access only to $\delta g_1^Z$.  Using \eq{heptosilh}, the  expected bounds on the HEP parameter translate onto  a  per-mille level constraint on $\delta g_1^Z$
\be
|\delta g_1^Z|\lesssim 0.001\,,
\ee
for the HL-LHC ($5\%$ systematics) and assuming a new physics scale above $3$ TeV.
Processes with other diboson final states, test complementary directions in the $\delta g_1^Z$ and $ \delta \kappa_\gamma-\hat S$ plane,  as we  illustrate in  the left panel of figure~\ref{cvsm2}. The colored lines indicate the directions along which the linear order new physics effects cancel, namely they are the approximate flat direction associated to the corresponding process. In particular, dashed lines correspond to parton-level processes $\bar q q\to W_LW_L/Z_Lh$,  as derived from table~\ref{Wilsons}, with polarized initial quarks. The solid red line corresponds to the approximate flat direction for the full $pp\to W_LW_L/Z_Lh$ process, obtained by weighting the interference terms between the polarized SM and BSM amplitudes by the corresponding parton luminosities.
Since the up and down luminosities ratio varies from $1.4$ to $1.6$ in the relevant energy range, this estimate is nearly independent of the center of mass energy.\footnote{We thank S. Gupta for pointing out a mistake in the first version of the figure.}

The gray shaded area in figure~\ref{cvsm2} shows bounds from LEP2~\cite{LEP:2003aa}. These bounds depend also on the parameter  $\lambda_\gamma$, which for simplicity we have taken to zero, a conservative choice in our comparison.
Our analysis is instead insensitive to (small values of) $\lambda_\gamma$, because of the non-interference rules discussed before. This comparison allows us to conclude  that, in the context of universal theories, LEP2  bounds will be order-of-magnitude improved by the HL-LHC, at least in the $\delta g_1^Z$ direction.
\\

In section~\ref{bsmsection}, we have further discussed explicit realizations of universal theories, which we can refer to as ``general SILH theories" and include e.g. theories with extra gauge bosons or extra-dimensions, holographic versions of composite Higgs or little Higgs models. In these theories $\delta \kappa_\gamma$ (and $\lambda_\gamma$)
 arise only at the one-loop level, and are therefore expected to be small. Similarly, for large $g_*$,  $W$ and $Y$ are small, see  for instance \eq{compopred}.
As a result, the only relevant parameters  are 
$\hat S$ and $\delta g_1^Z$, that can be induced at  tree-level. 
These parameters enter in the HEPs, \eq{heptosilh}, and provide then a strong motivation for our analysis. 
The results are  shown in the right panel of figure~\ref{cvsm2}. Present limits on $\hat S$ come from LEP measurements on the $Z$-pole, and  we do not expect that
the LHC will improve them any further (such an improvement  would  require very accurate measurements of the $W_LW_L/Z_Lh$ channels).

This result can be better appreciated in the specific context of composite Higgs models with $O(4)$ symmetry, where the two parameters are related according to \eq{compopred}, $\delta g_1^Z\simeq -\hat S/2c^2_{\theta_W}$ (corresponding to $c_B=c_W$), as shown by a blue solid line in the plot.
In this context it becomes remarkable that the size of the constraint on $\hat S$ from LEP (which is considered one of the most precise measurements of the EW sector) is comparable with
that on $\delta g_1^Z$, obtained from our analysis at the HL-LHC. 
Our bound will not only be competitive, but also complementary to LEP.
Indeed, the LEP measurement is affected by  a number of other low-energy effects. First of all, measurements of $\hat S$ are correlated experimentally with $\hat T$, as can be seen by the  grey bands in the right panel of figure~\ref{cvsm2}, corresponding to $\hat T=0$ and marginalization over $\hat T$, respectively. In addition, LEP has access  to the low-energy value of $\hat S$, which differs from the high-energy value (to which our analysis is sensitive) by renormalization effects induced by other operators~\cite{Barbieri:2007bh}.
\\

\begin{figure}[t]
\centering
\includegraphics[width=0.465\textwidth]{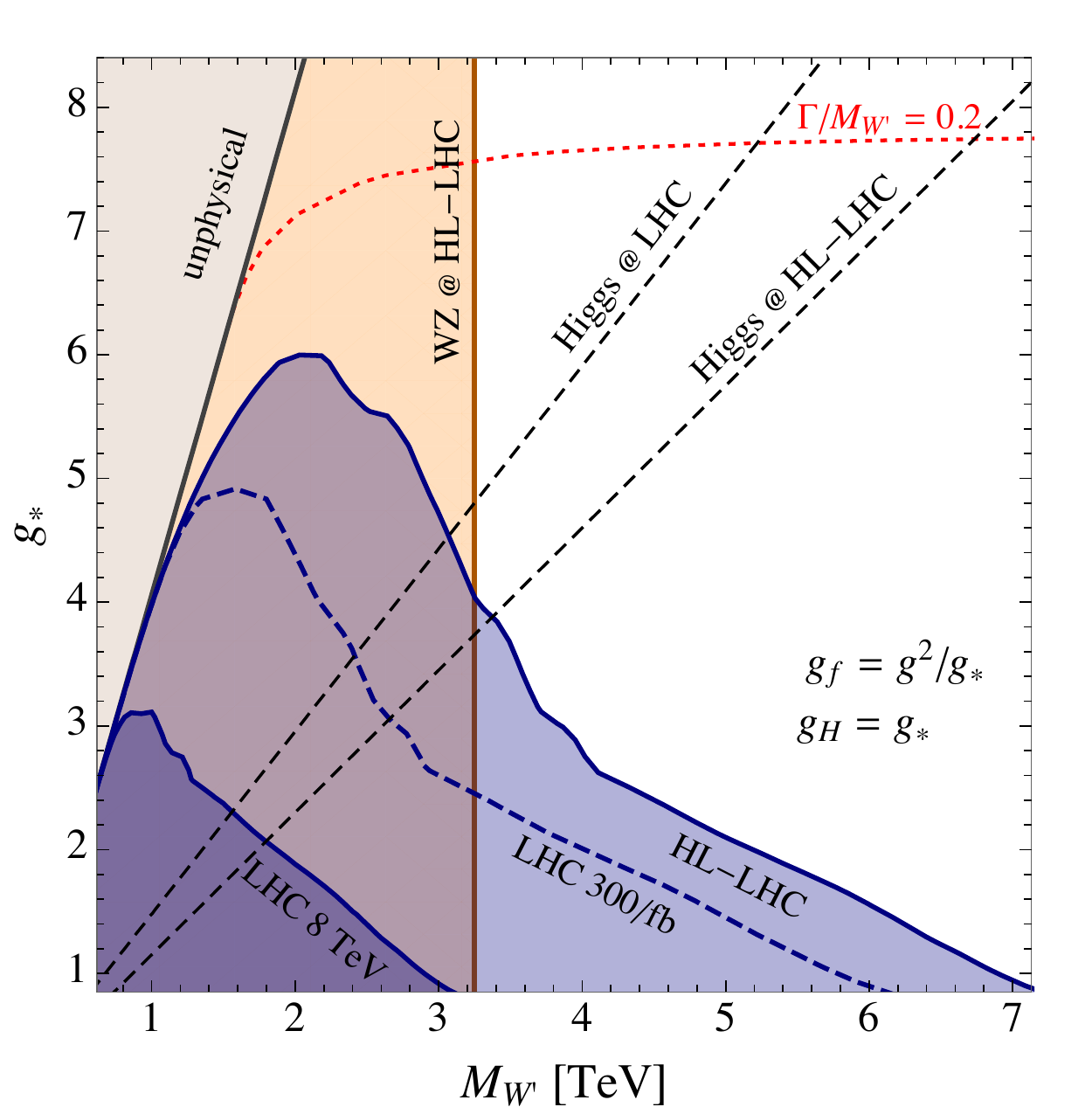}\ \ \
\includegraphics[width=0.48\textwidth]{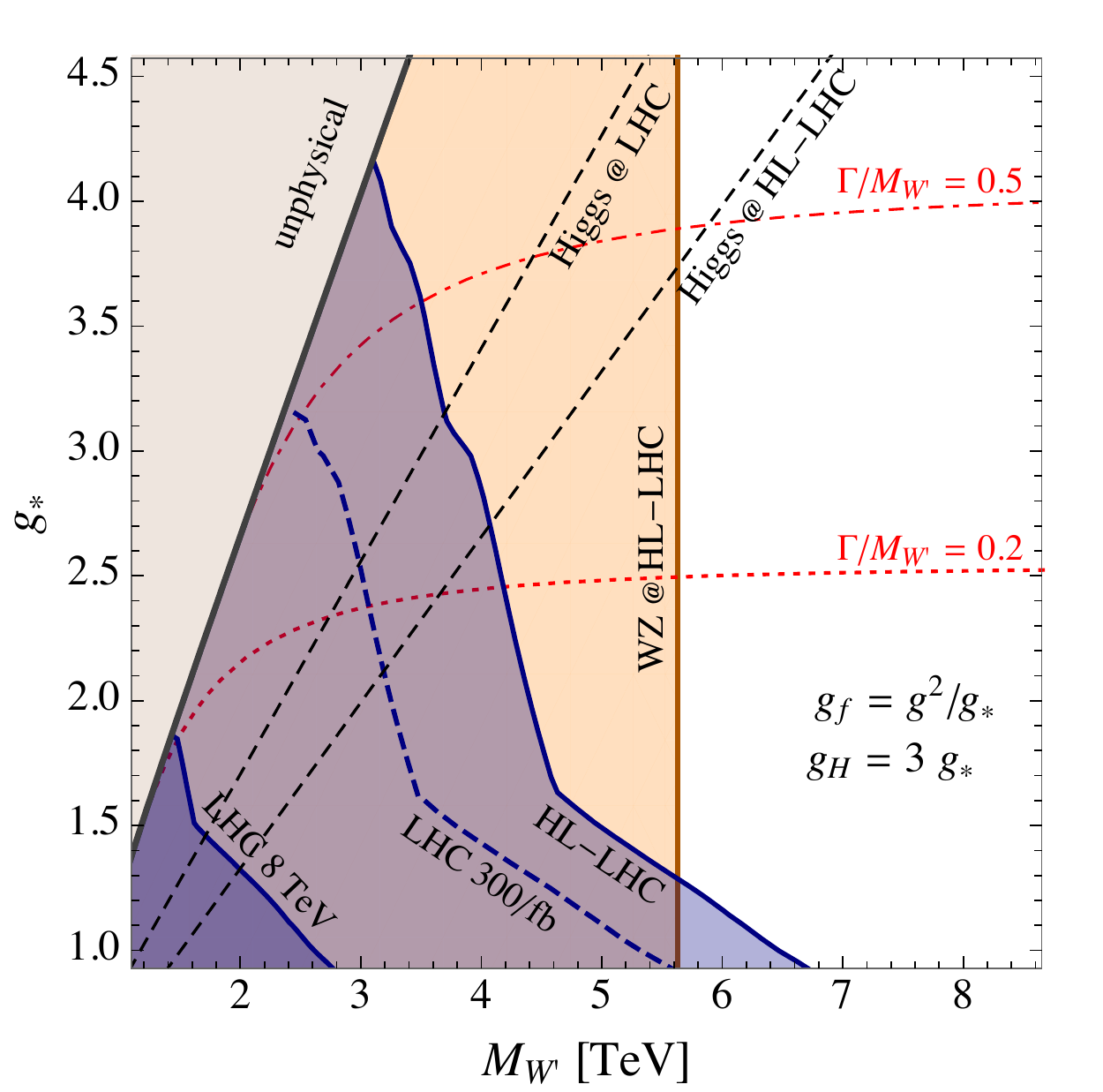}
\caption{Bounds on the mass and coupling of a  heavy triplet resonance
(see text).
}
\label{combounds}
\end{figure}

Our discussion so far has been based on the assumption that the new dynamics is much heavier than the LHC kinematic reach, so that an EFT approach is appropriate. It is however instructive to confront these indirect searches in the EFT framework, with direct resonance searches in explicit models. 
We do this in figure~\ref{combounds}
in the context of models with  heavy vector triplet resonances $W^\prime$, as introduced in \eq{hvt}. 
For concreteness,
we have performed this comparison with vector resonances arising from composite Higgs models,
 fixing the $W^\prime$ couplings according to the scaling described in model B of Ref.~\cite{Pappadopulo:2014qza}.
More specifically, in  \eq{hvt} we chose $g_H=g_*$ (left panel of figure~\ref{combounds}) and $g_H=3g_*$ (right panel), while   the coupling to  fermions  is  controlled by $g_f=g^2/g_*$, reflecting the fact that fermions are external to the strong dynamics.\footnote{In the notation of  \cite{Pappadopulo:2014qza}, we have    $g_\rho=g_*$, $c_H=g_H/g_*$,  and $c_F=1$.}
The region  excluded by our results in $WZ$ production is shown in orange (using \eq{Wpred}), while in
purple  is shown the exclusion from direct searches  at the LHC and HL-LHC~\cite{Thamm:2015zwa}.
The dashed red lines  show different values of  $\Gamma/M_{W'}$: in
regions where  $\Gamma/M_{W'}\gtrsim 0.2$  the  resonance becomes broad
and bounds from direct searches are inaccurate.
Dashed lines provide    bounds from Higgs physics.  In particular, regions above these lines
lead to  deviations from  the Higgs coupling to
$VV$   larger than $10\%$ (bound expected at the LHC) and $5\%$
(bound expected at the   HL-LHC).
Figure~\ref{combounds} shows that indirect bounds from our analysis
can be   stronger than those  from direct searches.
This is especially relevant for large couplings between $W'$  and the Higgs ($g_H\gg g_f$).

\section{Conclusions and Outlook}\label{sec:conc}

We identified a set of new physics indirect effects (dubbed ``High Energy Primary'', or HEP) that can be probed in high-energy diboson production at the LHC and at future colliders. These probes will be part of an extensive precision program to be performed at current and future hadronic machines. The four HEP parameters describe the most general BSM effects that grow quadratically with the energy and that interfere with the SM in measurements that are inclusive over the bosons decay products. For such measurements, and for BSM theories of the ``Weak'' type,\footnote{We stress once again that ``Weak'' refers here to the interaction of light quarks and transverse vector bosons. The longitudinals and the Higgs bosons might well be strongly-interacting. Composite Higgs models are indeed ``Weak'' theories in this context, and our strategy is perfectly suited to deal with them.}  in which the BSM contributions to the amplitudes do not exceed the SM one, the HEPs are the only effects that are visible in the high-energy diboson processes. Hence they form a complete basis that can be used for a global interpretation of these measurements. The HEP parameters map very simply to $d=6$ EFT Wilson coefficients.

We also estimated the reach of the leptonic $WZ$ process on the HEP $a^{(3)}_q$, showing that even the LHC run-$2$ dataset will be sufficient to start probing unexplored territories. The full LHC luminosity will improve the LEP reach for aTGC by one order of magnitude. 
By probing the HEP parameters at the HL-LHC, one will have comparable and complementary reach as LEP in new physics scenarios in which the LEP bounds on the HEPs come from the S parameter, which was much better constrained than the aTGC.
 We also showed that the indirect reach on the HEPs can be superior to the one of direct searches, even in BSM models where $s$-channel resonance production occurs in the same channel that is used to probe the HEPs. We stressed throughout the paper that our strategy crucially relies on {\emph{accuracy}} on both the experimental measurements and on the SM predictions. More careful studies would be needed in order to assess if and how the accuracy we assumed can be achieved. From the theoretical side we checked that the parton distribution function uncertainties are small and that scale variations are under control already at NLO in QCD. An assessment of the uncertainties at NNLO,  including NLO EW, would be needed.

We can then conclude that the measurement of the HEP parameter $a^{(3)}_q$, 
 together with the determination of the $W$ and $Y$ parameters in Drell-Yan processes studied in Ref.~\cite{Farina:2016rws}, provide at present the most promising precision tests of the EW sector to be performed at the LHC. Although the  projected limits on $W$ and $Y$ are very strong \cite{Farina:2016rws},  their expected size is very small in certain BSM such as the SILH (see \eq{compopred}). The HEP parameters are instead unsuppressed because they correspond to operators that involve the Higgs field. Hence they probe  directly the EW symmetry breaking sector, as shown in  section~\ref{bsmsection}. 
 These EWPT at the LHC can improve and complement those from LEP.

Our study should be extended in several directions. In the first place, other diboson processes should be studied in order to explore, following table~\ref{Wilsons}, all the HEP directions. The $WZ$ process we considered is indeed only sensitive to one of the four HEP parameters: $a_q^{(3)}$. Since $Wh$ production is also only sensitive to $a_q^{(3)}$, the most urgent channels to be explored are $WW$ and $Zh$, that however suffer from large backgrounds. Backgrounds come from transverse polarizations in the case of $WW$ and from jets faking a boosted Higgs in the case of $Zh$. Suitable strategies should be identified in order to deal with these backgrounds, including the study of differential distributions of the boson decay products and progresses in boosted Higgs reconstruction techniques. 

A careful study of differential distributions, with refined multivariate analysis techniques, might also improve the reach in the leptonic $WZ$ channel we considered in this paper. In this respect it is important to remark that we didn't explore this possibility because we designed our analysis having in mind a measurement of the $p_{T,V}$ differential cross-section, in a fiducial region, to be  eventually reinterpreted by a $\chi^2$ fit of the HEP parameters. This allowed us to parametrize systematical and theoretical errors simply in terms of the relative uncertainty parameter $\delta_{\textrm{syst}}$, but prevented us from exploiting fully differential informations. Doing so would require an experimental analysis that is more similar to a BSM search than to a SM measurement. The impact of systematical and theoretical uncertainties is much harder to quantify with this second approach. 

Our analysis needs improvement from the theoretical side as well. Our HEPs offer a complete parameterization of BSM effects only in measurements that are inclusive over the boson decay products angular distributions. 
Otherwise, and in particular if the azimuthal decay angle is measured, the interference among different helicity diboson amplitudes ``resurrects'' \cite{Panico:2017??} and there is no reason to restrict to BSM effects in the longitudinal diboson channels as we did in section~\ref{sec:thf}. It is straightforward to extend the HEP parameterization to transverse amplitudes, and furthermore we expect that it should be relatively easy to disentangle the new transverse HEP parameters from the longitudinal ones through the study of azimuthal distributions. This is left to future work.

Departing from the `Weak'' hypothesis one might also want to test  scenarios where BSM effects can overcome the SM amplitude. 
In this case, for a global analysis, effects that do not interfere with the SM should  be included. For instance, ${\overline{u}}_\pm d_{\mp}$--initiated production or  same-helicity quark anti-quark collisions mediated by dipole operators. 
We don't feel the need of such an extension at the current stage, both because of the limited BSM motivation of non--``Weak'' new physics and because these scenarios are most likely better probed in other channels (for instance, ``Remedios'' is probed in Drell--Yan) than dibosons.

\section*{Acknowledgements}
We are grateful to Celine Degrande, Senka Djuric and Olivier Mattelaer for support with the UFO model \texttt{EWdim6}, Admir Greljo for providing details of the analysis of Ref.~\cite{Falkowski:2016cxu}, and Andrea Thamm and Riccardo Torre for help with the direct searches constraints in Fig.~\ref{combounds}.
We also thank Sandeepan Gupta and Da Liu for collaboration in an earlier version of this project and Riccardo Rattazzi for discussions.
RF thanks CERN Theory Division for hospitality and support during the completion of this work. 
RF is supported by the ``Rita Levi-Montalcini'' program for young researches of MIUR. 
AP is  supported by the Catalan ICREA Academia Program. 
AP and GP are   supported by the grants FPA2014-55613-P, 2014-SGR-1450 and by the Severo Ochoa excellence program SO-2012-0234.
GP is also supported by the European Commission through the Marie Curie Career Integration Grant 631962 via the DESY-IFAE cooperation exchange.

\appendix

\section{Amplitude decomposition}\label{appA}

The particles involved in high-energy diboson production are the quarks and anti-quarks doublets and singlets and the Higgs doublet, which groups together the Higgs particles and the Goldstone boson states $|w^\pm\rangle$ and  $|z\rangle$ associated with longitudinally-polarised vector bosons. In terms of physical particles, the Higgs doublet particle multiplet $|\Phi\rangle_i$ and the anti-particle one $|{\overline{\Phi}}\rangle^i$ reads
\begin{equation}\label{uno}
\displaystyle
|\Phi\rangle_i=\left[\begin{array}{c}\displaystyle
|w^+\rangle\\
\displaystyle
\frac1{\sqrt{2}}(|h\rangle-|z\rangle)
\end{array}
\right]_i\in \;{\mathbf{2}}_{\mathbf{1/2}}\,,\;\;\;\;\;
|{\overline{\Phi}}\rangle^i=\left[\begin{array}{c}\displaystyle
-|w^-\rangle\\
\displaystyle
\frac1{\sqrt{2}}(|h\rangle+|z\rangle)
\end{array}
\right]_i\in \;{\mathbf{\overline{2}}}_{\mathbf{-1/2}}\,,
\end{equation}
while for the quark anti-quark multiplets we have
\begin{eqnarray}\label{due}
&&\displaystyle
|q_-\rangle_i=\left[\begin{array}{c}\displaystyle
|u_-\rangle\\
\displaystyle
|d_-\rangle
\end{array}
\right]_i\in \;{\mathbf{2}}_{\mathbf{1/6}}\,,\;\;\;\;\;
|u_+\rangle\in \;{\mathbf{1}}_{\mathbf{2/3}}\,,\;\;\;\;\;
|d_+\rangle\in \;{\mathbf{1}}_{\mathbf{-1/3}}\,,
\nonumber\\
&&\displaystyle
|{\overline{q}}_+\rangle^i=\left[\begin{array}{c}\displaystyle
|{\overline{u}}_+\rangle\\
\displaystyle
|{\overline{d}}_+\rangle
\end{array}
\right]_i\in \;{\mathbf{\overline{2}}}_{-\mathbf{1/6}}\,,\;\;\;\;\;
|{\overline{u}}_-\rangle\in \;{\mathbf{1}}_{\mathbf{-2/3}}\,,\;\;\;\;\;
|{\overline{d}}_-\rangle\in \;{\mathbf{1}}_{\mathbf{1/3}}\,.
\nonumber\\
\end{eqnarray}
Eq.~\ref{uno} requires some clarification. It is obtained from the standard expression for the Higgs doublet field $\Phi=(-i\varphi_+,(h+i\varphi_0)/\sqrt{2})$ by quantising the Goldstone fields using a creation/annihilation operators decomposition that contains unconventional $i$ factors. Equivalently, it can be obtained from the standard decomposition by reabsorbing a $-i$ factor in the Goldstone particles states. This automatically keeps track of the $-i$  factor that appears in the Equivalence Theorem relation \cite{Chanowitz:1985hj,Wulzer:2013mza} among longitudinal vectors and Goldstone boson external states.

Scattering amplitudes involving these particles as external states transform as tensors under the $G_{\rm{SM}}$ group, and the $G_{\rm{SM}}$ invariance of theory ensures that they must be invariant tensors. The tensor structure is particularly simple for $u_+{\overline{u}}_-$ and $d_+{\overline{d}}_-$ initial states since only two indices are those from the Higgs doublets, namely the amplitudes have the form
\begin{equation}
(A_u)_{i}^{j}=\langle{\overline{\Phi}}_i\,\Phi^j | T |u_+{\overline{u}}_-\rangle\,,\;\;\;\;\;(A_d)_{i}^{j}=\langle{\overline{\Phi}}_i\,\Phi^j| T |d_+{\overline{d}}_-\rangle\,,
\end{equation}
where $T$ denotes the $T$-matrix. There is of course only one invariant tensor with one ${\mathbf{2}}$ and one ${\mathbf{\overline{2}}}$ index, namely $\delta^i_j$, therefore 
\begin{equation}\label{tre}
(A_u)_{i}^{j}=-a_u \delta^j_i\,,\;\;\;\;\;(A_d)_{i}^{j}=-a_d \delta^j_i\,.
\end{equation}
The case of $q_-{\overline{q}}_+$ initial state is a bit more complicated because the amplitude has $4$ indices
\begin{equation}\label{quattro}
(A_q)_{ik}^{jl}=\langle{\overline{\Phi}}_i\,\Phi^j| T |(q_-)_k({\overline{q}}_+)^l\rangle\,.
\end{equation}
A total of two invariants are present in the tensor product of two doublets and two anti-doublets. They correspond to combining the Higgs doublets indices to form either a singlet or a triplet, and next contracting them with the appropriate combination of the fermion doublets. The amplitude decomposition thus reads
\begin{equation}
(A_q)_{ik}^{jl}=a_q^{(1)}\delta^j_i\delta^l_k+
a_q^{(3)}(\sigma^\alpha)_{i}^{\;\;j}(\sigma^\alpha)_{k}^{\;\;l}
\,,
\end{equation}
where $\sigma$ are the Pauli matrices and the sum over $\alpha$ is understood.

Up to now we only considered ${\overline{\Phi}}\Phi$ final state amplitudes. Those involving $\Phi\Phi$ or ${\overline{\Phi}}{\overline{\Phi}}$ final states trivially vanish, being forbidden by Hypercharge conservation (i.e., by the need of forming an invariant tensor under ${\textrm{U}}(1)_Y$) for same-quark-flavour initial states. This results in a number of constraints 
\begin{eqnarray}\label{cinque}
&&\langle{{\Phi}}^i\,\Phi^j | T |u_+{\overline{u}}_-\rangle=0\,,\;\;\;\;\;\langle{\overline{\Phi}}_i\,\overline{\Phi}_j | T |u_+{\overline{u}}_-\rangle=0\,,\nonumber\\
&&\langle{{\Phi}}^i\,\Phi^j | T |d_+{\overline{d}}_-\rangle=0\,,\;\;\;\;\;\langle{\overline{\Phi}}_i\,\overline{\Phi}_j | T |d_+{\overline{d}}_-\rangle=0\,,\nonumber\\
&&\langle{{\Phi}}^i\,\Phi^j| T |(q_-)_k({\overline{q}}_+)^l\rangle=0\,,\;\;\;\;\;\langle{\overline{\Phi}}_i\,\overline{\Phi}_j| T |(q_-)_k({\overline{q}}_+)^l\rangle=0\,,
\end{eqnarray}
that are essential in order to obtain the final result.

By substituting eqs.~(\ref{uno}), (\ref{due}) in eqs.~(\ref{tre}), (\ref{quattro}) and (\ref{cinque}), the physical scattering amplitudes are easily expressed in terms of the $4$ amplitude coefficients $a_{u}$, $a_{d}$, $a_q^{(1)}$ and $a_q^{(3)}$, obtaining the results in eq.~(\ref{primaries0}). One important point must be taken into account when performing the substitution, related with the fact that in the main text we are only interested in scattering processes that occur in the $J=1$ angular momentum configuration. The momentum-space wave-function of the states is odd under the exchange of the boson momenta,\footnote{The reader might find this confusing if looking at eq.~(\ref{amp0}), which is even and not odd under $\cos\theta\rightarrow-\cos\theta$. However momenta exchange also entails the operation $\phi\rightarrow\phi+\pi$ on the azimuthal angle, which has been set to zero in eq.~(\ref{amp0}). The Jacob-Wick formula \cite{Jacob:1959at} foresees the dependence on $\phi$ to be $e^{\pm i\phi}$, making indeed the amplitude odd under momenta exchange.}  therefore $|hh\rangle$ and $|zz\rangle$ final states vanish by Bose symmetry and $|zh\rangle=-|hz\rangle$.

Notice that the exact same decomposition holds for the SM amplitudes, which are also in the $J=1$ eigenstate if the Yukawa couplings are negligible. The only difference is that of course the SM does not grow with the energy, hence there is no $E^2/4$ factor in \eq{amp0}. Explicitly, the SM amplitude coefficients are
\begin{equation}
\alpha_u=\frac{g_1^2}3\,,\;\;\;\;\;\alpha_d=-\frac{g_1^2}6\,,\;\;\;\;\;\alpha_q^{(1)}=\frac{g_1^2}{12}\,,\;\;\;\;\;\alpha_q^{(3)}=\frac{g_2^2}{4}\,.
\end{equation}


\end{document}